%% file: main.tex
\documentclass[a4paper]{article}
\pdfoutput=1 

\usepackage[utf8]{inputenc}
\usepackage[english]{babel}
\usepackage{amsmath,amsbsy,bm,amssymb,mathtools,dsfont,etoolbox,braket}
\usepackage{xspace}
\usepackage{geometry}
\usepackage{graphicx,float,placeins}
\usepackage[font=small,labelfont=bf]{caption}
\usepackage[subrefformat=parens,labelformat=parens]{subcaption}
\usepackage[bf,sf,pagestyles]{titlesec}
\usepackage{titling,abstract}
\usepackage[dvipsnames]{xcolor}
\usepackage{tikz,pgfplots,pgfplotstable}
\usepackage[mode=buildnew]{standalone}
\usepackage{multirow,makecell,array,tabularx,booktabs}
\usepackage{algorithm,algpseudocode}
\usepackage[title]{appendix}
\usepackage{siunitx}
\usepackage{csquotes}
\usepackage[nolist]{acronym}
\usepackage[bibstyle=numeric,citestyle=numeric-comp,sorting=none,backend=biber]{biblatex}
\usepackage{orcidlink}
\usepackage{hyperref}
\usepackage{cleveref}

\input{common}

\crefname{figure}{Fig.}{Figs.}
\Crefname{figure}{Fig.}{Figs.}
\crefname{table}{Tab.}{Tabs.}
\Crefname{table}{Tab.}{Tabs.}
\crefname{equation}{Eq.}{Eqs.}
\Crefname{equation}{Eq.}{Eqs.}
\crefname{section}{Sec.}{Secs.}
\Crefname{section}{Sec.}{Secs.}
\crefname{subsection}{Sec.}{Secs.}
\Crefname{subsection}{Sec.}{Secs.}
\crefname{algorithm}{Alg.}{Algs.}
\Crefname{algorithm}{Alg.}{Algs.}
\crefname{appendix}{App.}{App.}
\Crefname{appendix}{App.}{App.}

\pretitle{\begin{center}\sffamily\LARGE}
\preauthor{\begin{center}%
\large\sffamily \lineskip 0.5em%
\begin{tabular}[t]{c}}
\predate{\begin{center}\sffamily\large}

\newcommand{\email}[1]{\href{mailto:#1}{#1}}

\def\scaleA{1.}
\def\scaleB{1.}
\def\scaleC{1.}

\newcommand{\appendixsection}[1]{\subsection{#1}}

\hypersetup{%hidelinks,
colorlinks, linkcolor={blue!45!black},citecolor={blue!45!black},urlcolor={blue!45!black}}

\title{Explaining Quantum Circuits with Shapley Values: \\ Towards Explainable Quantum Machine Learning}
\author{\normalsize%
Raoul Heese\textsuperscript{1,*}\orcidlink{0000-0001-7479-3339}, 
Thore Gerlach\textsuperscript{2}\orcidlink{0000-0001-7726-1848},
Sascha Mücke\textsuperscript{3}\orcidlink{0000-0001-8332-6169},
Sabine Müller\textsuperscript{1}\orcidlink{0000-0003-1682-4576},
\\[.1cm]\normalsize
Matthias Jakobs\textsuperscript{3}\orcidlink{0000-0003-4607-8957}, and
Nico Piatkowski\textsuperscript{2}\orcidlink{0000-0002-6334-8042}
}
\date{\normalsize%
\textsuperscript{1}Fraunhofer ITWM, \textsuperscript{2}Fraunhofer IAIS, \textsuperscript{3}TU Dortmund
\\[.1cm]\normalsize
\textsuperscript{*}\email{raoul.heese@gmail.com}
}

\begin{document}

\maketitle

\begin{abstract}
\input{abstract}
\end{abstract}

\input{introduction}
\input{background}

\input{method}

\input{experiments}
\input{conclusions}

\section*{Acknowledgments}
\input{acknowledgments}

\FloatBarrier
\printbibliography[heading=bibintoc]

\appendix
\section*{Appendices}\phantomsection\addcontentsline{toc}{section}{Appendices}
\renewcommand{\thesubsection}{\Alph{subsection}}
\input{appendix}

\end{document}

%% file: common.tex
\usepackage{amsmath,amsbsy,bm,amssymb,mathtools,dsfont,etoolbox,braket}
\usepackage[nolist]{acronym}
\usepackage{siunitx}
\usepackage{xspace}
\usepackage{tikz}
\usepackage{pgfplots}
\usepackage{pgfplotstable}
\pgfplotsset{compat=newest}
\usetikzlibrary{backgrounds}
\pgfdeclarelayer{background}
\pgfdeclarelayer{foreground}
\usepgfplotslibrary{groupplots}
\usepgfplotslibrary{colormaps}
\pgfsetlayers{background,main,foreground}
\usetikzlibrary{calc,external}
\usetikzlibrary{shapes.misc}
\usetikzlibrary{patterns}
\usetikzlibrary{decorations.pathreplacing}
\usetikzlibrary{arrows,arrows.meta}
\usetikzlibrary{quantikz}
\usepgfplotslibrary{fillbetween}
\usepgfplotslibrary{statistics}
\usepackage{cleveref}

\makeatletter
\AtBeginDocument
{
	\def\ltx@label#1{\cref@label{#1}}%add braces
	\def\label@in@display@noarg#1{\cref@old@label@in@display{#1}}%remove braces
	\def\label@in@mmeasure@noarg#1{%
		\begingroup%
		\measuring@false%
		\cref@old@label@in@display{#1}%remove braces for multline, see https://tex.stackexchange.com/q/737204/2388
		\endgroup}%  
} %
\makeatother

\tikzset{cross/.style={cross out, draw=black, minimum size=2*(#1-\pgflinewidth), inner sep=0pt, outer sep=0pt},cross/.default={.15cm}}
\tikzset{}
\makeatletter
\pgfplotsset{
	boxplot prepared from table/.code={
		\def\tikz@plot@handler{\pgfplotsplothandlerboxplotprepared}%
		\pgfplotsset{
			/pgfplots/boxplot prepared from table/.cd,
			#1,
		}
	},
	/pgfplots/boxplot prepared from table/.cd,
	table/.code={\pgfplotstablecopy{#1}\to\boxplot@datatable},
	row/.initial=0,
	make style readable from table/.style={
		#1/.code={
			\pgfplotstablegetelem{\pgfkeysvalueof{/pgfplots/boxplot prepared from table/row}}{##1}\of\boxplot@datatable
			\pgfplotsset{boxplot/#1/.expand once={\pgfplotsretval}}
		}
	},
	make style readable from table=lower whisker,
	make style readable from table=upper whisker,
	make style readable from table=lower quartile,
	make style readable from table=upper quartile,
	make style readable from table=median,
	make style readable from table=lower notch,
	make style readable from table=upper notch
}
\makeatother

\definecolor{quantumviolet}{HTML}{53257F}
\definecolor{quantumgray}{HTML}{555555}
\colorlet{colorXfirst}{quantumgray}
\colorlet{colorAfirst}{quantumviolet}
\definecolor{colorAsecond}{HTML}{5B4B8A}
\definecolor{colorAthird}{HTML}{666699}
\definecolor{colorBfirst}{HTML}{774360}
\definecolor{colorBsecond}{HTML}{B25068}
\definecolor{colorBthird}{HTML}{E7AB79}
\definecolor{colorCfirst}{HTML}{2F5D62}
\definecolor{colorCsecond}{HTML}{5E8B7E}
\definecolor{colorCthird}{HTML}{A7C4BC}
\colorlet{colorXa}{colorXfirst}
\colorlet{colorXafill}{colorXa!20!white}
\colorlet{colorXb}{colorXfirst!20!white}
\colorlet{colorXbfill}{colorXb!20!white}
\colorlet{colorXc}{colorXfirst!40!white}
\colorlet{colorXcfill}{colorXc!20!white}
\colorlet{colorAa}{colorAfirst}
\colorlet{colorAafill}{colorAa!20!white}
\colorlet{colorAb}{colorAsecond}
\colorlet{colorAbfill}{colorAb!20!white}
\colorlet{colorAc}{colorAthird}
\colorlet{colorAcfill}{colorAc!20!white}
\colorlet{colorBa}{colorBfirst}
\colorlet{colorBafill}{colorBa!20!white}
\colorlet{colorBb}{colorBsecond}
\colorlet{colorBbfill}{colorBb!20!white}
\colorlet{colorBc}{colorBthird}
\colorlet{colorBcfill}{colorBc!20!white}
\colorlet{colorCa}{colorCfirst}
\colorlet{colorCafill}{colorCa!20!white}
\colorlet{colorCb}{colorCsecond}
\colorlet{colorCbfill}{colorCb!20!white}
\colorlet{colorCc}{colorCthird}
\colorlet{colorCcfill}{colorCc!20!white}

\newcommand{\HADAMARD}{\mathrm{H}}
\newcommand{\NOT}{\mathrm{X}}
\newcommand{\RY}{\mathrm{R}_{\mathrm{Y}}}
\newcommand{\RZ}{\mathrm{R}_{\mathrm{Z}}}
\newcommand{\PHASE}{\mathrm{P}}
\newcommand{\SX}{\sqrt{\mathrm{X}}}
\newcommand{\CNOT}{\mathrm{CX}}
\newcommand{\SWAP}{\mathrm{S}}
\newcommand{\CPHASE}{\mathrm{CP}}
\newcommand{\QCstatevector}{\texttt{state\_CPU}\@\xspace}
\newcommand{\QCaer}{\texttt{shot\_CPU}\@\xspace}
\newcommand{\QCehningen}{\texttt{ibmq\_ehningen}\@\xspace}
\newcommand{\QCoslo}{\texttt{ibm\_oslo}\@\xspace}
\newcommand{\includefigure}[2]{%
\includegraphics[scale=#1]{#2.pdf}%
}

\newcommand{\const}{\operatorname{const.}}
\newcommand{\ind}{\mathbb{I}}
\newcommand{\expect}[1]{\mathbb{E}\!\left[#1\right]}
\newcommand{\std}[1]{\sigma\!\left[#1\right]}
\newcommand{\RV}[1]{\bm{#1}} % random variables
\newcommand{\VEC}[1]{\vec{#1}} % vectors
\newcommand{\RVEC}[1]{\VEC{\RV{#1}}} % random vectors
\newcommand{\subS}{S} % S subset 
\newcommand{\allS}{\mathfrak{S}} % S total set

\newcommand{\ie}{i.e.\@\xspace}
\newcommand{\eg}{e.g.\@\xspace}

\newcommand{\iid}{i.i.d.\@\xspace}
\newcommand{\wrt}{w.r.t.\@\xspace}

\begin{acronym}	
	\acro{ML}[ML]{machine learning}
	\acro{AI}[AI]{artificial intelligence}
	\acro{QML}[QML]{quantum machine learning}	
	\acro{XAI}[XAI]{explainable \ac{AI}} %\acro{XML}[XML]{explainable \ac{ML}}
	\acro{XQML}[XQML]{\emph{explainable \ac{QML}}}	
	\acro{SV}[SV]{Shapley value}
	\acro{QSV}[SVQX]{\emph{Shapley value for quantum circuit explanations}} %\acro{QSV}[QSV]{\emph{Quantum-applied Shapley value}}
	\acrodefplural{QSV}{\emph{Shapley values for quantum circuit explanations}} %\acrodefplural{QSV}{\emph{Quantum-applied Shapley values}}
	\acro{QSVold}[QSV]{\emph{quantum Shapley value}}
	\acrodefplural{QSVold}{\emph{quantum Shapley values}}
	\acro{CPU}[CPU]{central processing unit}		
	\acro{GPU}[GPU]{graphics processing unit}
	\acro{RAM}[RAM]{random-access memory}
	\acro{QPU}[QPU]{quantum processing unit}
	\acro{NISQ}[NISQ]{noisy intermediate-scale quantum}
	\acro{POVM}[POVM]{positive operator-valued measure}
	\acro{SPAM}[SPAM]{state preparation and measurement}
	\acro{MEM}[MEM]{measurement error mitigation}	
	\acro{VQA}[VQA]{variational quantum algorithm}
	\acro{VQE}[VQE]{variational quantum eigensolver}
	\acro{SVM}[SVM]{support vector machine}
	\acro{QSVM}[QSVM]{quantum support vector machine}
	\acro{QNN}[QNN]{quantum neural network}
	\acro{QAOA}[QAOA]{Quantum Approximate Optimization Algorithm}
	\acro{QGAN}[QGAN]{quantum generative adversarial network}
	\acro{QFT}[QFT]{quantum Fourier transform}
	\acro{QT}[QTree]{quantum decision tree classifier}
	\acro{NAS}[NAS]{\emph{neural architecture search}}
	\acro{AutoML}[AutoML]{automatic machine learning}
	\acro{AutoQML}[AutoQML]{\emph{automatic quantum machine learning}}	
	\acro{QCAS}[QAS]{\emph{quantum circuit architecture search}}
\end{acronym}

%% file: abstract.tex
Methods of artificial intelligence (AI) and especially machine learning (ML) have been growing ever more complex, and at the same time have more and more impact on people's lives. This leads to explainable AI (XAI) manifesting itself as an important research field that helps humans to better comprehend ML systems. In parallel, quantum machine learning (QML) is emerging with the ongoing improvement of quantum computing hardware combined with its increasing availability via cloud services. QML enables quantum-enhanced ML in which quantum mechanics is exploited to facilitate ML tasks, typically in the form of quantum-classical hybrid algorithms that combine quantum and classical resources. Quantum gates constitute the building blocks of gate-based quantum hardware and form circuits that can be used for quantum computations. For QML applications, quantum circuits are typically parameterized and their parameters are optimized classically such that a suitably defined objective function is minimized. Inspired by XAI, we raise the question of the explainability of such circuits by quantifying the importance of (groups of) gates for specific goals. To this end, we apply the well-established concept of Shapley values. The resulting attributions can be interpreted as explanations for why a specific circuit works well for a given task, improving the understanding of how to construct parameterized (or variational) quantum circuits, and fostering their human interpretability in general. An experimental evaluation on simulators and two superconducting quantum hardware devices demonstrates the benefits of the proposed framework for classification, generative modeling, transpilation, and optimization. Furthermore, our results shed some light on the role of specific gates in popular QML approaches.

%% file: introduction.tex
\section{Introduction} \label{sec:introduction}
\Ac{ML} has a significant impact on many applications in different domains.
With the ongoing improvement of \ac{ML} models and their growing complexity, another aspect is becoming increasingly important in addition to pure performance: the explainability of \ac{ML} systems \cite{molnar2022}.
Explainability\footnote{We do not distinguish between \emph{explainability} and \emph{interpretability} because these terms are not strictly defined and are often used interchangeably in the \ac{ML} community \cite{marcinkevics2020}.} refers to methods that make the behavior of \ac{ML} systems---or, more generally, \ac{AI} systems \cite{mukhamediev2022}---comprehensible for humans.
Realizing \ac{XAI} is a highly non-trivial task with a potentially great impact on many applications and can therefore be considered as an important research field.
For example, \ac{XAI} can be used to analyze the fairness of models to avoid discrimination, or their security against adversarial attacks, to name just two important aspects.
Currently, reasoning about decisions is typically only possible to a very limited extent or even intractable for state-of-the-art \ac{ML} models.
A review of \ac{XAI} exceeds the scope of this paper and we instead refer to, \eg, \cite{marcinkevics2020,linardatos2021,minh2022,molnar2022,saeed2023} and references therein.\par
Apart from explainability, another promising research direction of \ac{ML} that has recently arisen with the emergence of new technology is \ac{QML} \cite{biamonte2017,schuld2018}.
This field addresses how quantum computers (or quantum information processing in general \cite{nielsen2011}) can be used to provide a benefit for \ac{ML}.
To identify possible advantages, the structure of the data seems to be key \cite{huang2021,liu2021,caro2022}.
However, since currently only \ac{NISQ} devices \cite{preskill2018} with limited capabilities are available, a practical application of \ac{QML} is restricted to toy examples.
On the other hand, the identification of a quantum advantage is not necessarily of central importance for current research \cite{schuld2022}.
Instead, the study of fundamental aspects of \ac{QML} and their prospective exploitation with better hardware in the future might be more promising.\par
With this in mind, it seems natural to study explainability for \ac{QML} in a similar way as for \emph{classical} (\ie, non-quantum) \ac{ML}, giving rise to \ac{XQML} with the goal to provide humanly understandable interpretations of \ac{QML} systems.
This aspect of \ac{QML} has been virtually unexplored at the time of the initial submission of this manuscript, except for one other study in the context of malware detection \cite{mercaldo2022}.
In the meantime, however, there have been a number of new research efforts that indicate a growing awareness of the topic. For example, in the form of review articles \cite{power2024,gujji2024} that also introduce novel methods \cite{gilfoster2024} as well as articles focusing on quantum neural networks \cite{ciaramella2023,jinkai2024,pira2024}, the visualization of model behavior \cite{shaolun2023,lin2024}, or reinforcement learning \cite{cuellar2024,flamini2024}, just to name a few.
\par
\ac{XAI} methods can be divided into model-specific approaches and model-agnostic approaches.
While the former are limited to specific model classes, the latter can be used for arbitrary models.
Hence, model-agnostic \ac{XAI} methods can be applied to \ac{QML} models in the same way as to classical \ac{ML} models.
This leaves us with a rich set of classical tools that can be directly explored and potentially refined for \ac{XQML} \cite{power2024,gilfoster2024}.
Since \ac{QML} is still in a very early stage in comparison to \ac{ML}, explainability can be adopted from the beginning for comparatively simple tasks that are much easier to analyze than the classical state-of-the-art problems. Their exploration could lead to fundamental insights, similar to the classical analog.\par
In the study of the decision making process of classical \ac{ML} systems, two central, interrelated questions are typically addressed by \ac{XAI}:
\begin{enumerate}
	\item What is the influence of the model inputs (\ie, features) on its predictions?
	\item How does the model architecture (its components and/or trainable parameters) affect its predictions?
\end{enumerate}
The first question takes an external perspective by investigating the (potentially black-box) relationship between inputs and outputs, and can be considered the more commonly asked. The second shifts attention inward, exploring how the architecture and parameters of the model itself affect its behavior.\par
In the present paper, we limit ourselves to the second perspective and propose a framework for approaching model architecture investigations in an \ac{XQML} context. Specifically, our framework allows to explore the influence of gates (or groups of gates) as the building blocks of gate-based quantum computations using \acp{SV} \cite{shapley1953,aumann1974}, a well-known model-agnostic tool in \ac{XAI} \cite{strumbelj2010,lundberg2017,sundararajan2017}. While our considerations are generally applicable to any kind of gate-based quantum computation, we focus mostly on \ac{QML} applications driven by our original motivation for such an analysis. We refer to our approach as \acp{QSV}.\par
The study of internal structures of \ac{ML} models is an important topic in deep learning, for example to visualize the behavior of complex systems \cite{zhang2018}. Often, it is closely related to the search for an optimal model design, especially in the context of neural networks, where it is known as \ac{NAS} \cite{salehin2024,salmanipouravval2025} and can also be based on \acp{SV} \cite{leon2015,stier2018,xiao2022,diazortiz2023,oloulade2025}. \Ac{NAS} is a special branch of \ac{AutoML}, a research field concerned with the automation of the entire \ac{ML} pipeline \cite{barbudo2023}. This concept has also been transferred to \ac{QML}, where it is known as \ac{AutoQML} \cite{klau2023,rybotycki2024}. In this connection, several approaches have been proposed to optimize quantum circuit designs for \ac{QML} applications in the sense of a \ac{QCAS} \cite{ding2022,du2022b,franken2022,giovagnoli2023,chen2024}. While our proposed \acp{QSV} can in principle also be used in the context of \ac{QCAS}, we focus on the explainability aspect in this paper and demonstrate its potential for circuit design optimization only by way of example.\par
Our main contributions can be listed as follows:
\begin{itemize}
	\item We present \acp{QSV} as a method to measure the impact of every gate in a (variational) quantum circuit on the overall expressibility, entanglement capability, or any arbitrary quantity of interest.
	\item The proposed method is grounded in the theory of uncertain \acp{SV} and therefore extends this domain by a quantum approach while facilitating inherent robustness to the noise from NISQ devices.
	\item \acp{QSV} can be leveraged for \ac{XQML}, which is the primary focus in this work.
	\item We perform experiments on simulators and actual quantum hardware to showcase different use cases.	
	\item A software toolbox and the data from our experiments are publicly available online.
\end{itemize}
The remainder of this manuscript is divided into four sections.
In \cref{sec:background}, we introduce the prerequisites for our proposed method, which we present in \cref{sec:method}.
We subsequently study experimentally in \cref{sec:experiments} how our method can be used for a variety of applications.
Finally, we conclude with a summary and outlook in \cref{sec:conclusions}.

%% file: background.tex
\section{Background} \label{sec:background}
In the present section, we outline the prerequisites for our proposed \ac{XQML} method of \acp{QSV}.
For this purpose, we first present a short summary of \acp{SV} and how they can be applied to classical \ac{ML}.
Next, we give a brief introduction to \ac{QML}.

\subsection{Shapley values}
\Acp{SV} \cite{shapley1953,aumann1974} are a concept originating in game theory that can be used to evaluate the performance of individual players in a coalition game.
By definition, they fulfill many desirable properties that distinguish them as a particularly suitable evaluation measure for this purpose.
Presume a group of $N \in \mathbb{N}$ players denoted by the set $\allS := \{1,\dots,N\}$, a coalition game is defined by the value function (or characteristic function)
\begin{equation} \label{eqn:v:regular}
	v : \mathcal{P}(\allS) \rightarrow \mathbb{R},
\end{equation}
that assigns a (real-valued) payoff $v(\subS)$ to each subset (or \emph{coalition}) of players $\subS \subseteq \allS$.
The \ac{SV} of player $i \in \allS$ is then given by the expectation value
\begin{equation} \label{eqn:phi:regular}
	\phi_i := \expect{ \RV{V}_i }
\end{equation}
of the random variable\footnote{All random variables are in bold.} $\RV{V}_i \sim \mathbb{V}_i$ with probability mass function
\begin{equation} \label{eqn:deltav:p:regular}
	\mathbb{V}_i(v) := \sum_{\subS \subseteq \allS \setminus \{i\}} 
	w(\subS) \cdot\ind(i, \subS, v).
\end{equation}
Here, we have introduced the weight
\begin{equation} \label{eqn:w}
	w(\subS) := \frac{1}{N} {N-1 \choose \vert \subS \vert}^{-1} = \frac{\vert 
		\subS \vert ! \left( N  - \vert \subS \vert - 1 \right) !}{N !}
\end{equation}
of coalition $\subS$ based on its set cardinality $\vert \subS \vert$.
Furthermore,
\begin{equation}
	\ind(i, \subS, v) := \begin{cases} 1, & \text{ if $\Delta_i v(\subS)=v$} \\ 
		0, 
		& \text{ otherwise}\end{cases}
\end{equation}
denotes the indicator function with the marginal contribution
\begin{equation} \label{eqn:deltav}
	\Delta_i v (\subS) := v(\subS \cup \{i\}) - v(\subS)
\end{equation}
of player $i$ to the coalition $\subS$.
In total, there are $2^N$ different coalitions including the grand coalition $\allS$ and the empty coalition $\varnothing$.\par
The \ac{SV} framework is very general with typical applications in economy and finance \cite{aumann1994,winter2002} as well as \ac{XAI} \cite{strumbelj2010}.
Due to the computational complexity of exact calculations, random sampling strategies can be employed to obtain approximate results \cite{fatima2007}.
A quantum algorithm for the calculation of (classical) \acp{SV} in an \ac{XAI} context was recently proposed in \cite{burge2024}.\par
In our subsequent analyses, we apply \acp{SV} to quantum circuits.
For this purpose, gates represent players and the value function is obtained from measurement results, for example to quantify the contribution of each gate to produce a desired distribution of bits.
Due to the universality of the theory, the value function can be chosen in many different ways depending on the specific use case.
Our approach is explained in more detail in \cref{sec:method}, where we also give detailed examples for possible value functions.

\subsection{Shapley values under uncertainty}
So far, we have considered a deterministic value function $v(\subS)$, \cref{eqn:v}.
However, for value functions which are based on \ac{QPU} calculations, such a deterministic behavior is not given since results from QPUs are subject to
\begin{itemize}
	\item the statistical uncertainty of measurements that results from the inherent uncertainty of quantum physics and the limited number of shots,
	\item the typically non-deterministic mapping of logical qubits onto physical qubits together with a decomposition of the circuit unitary that can be realized on the hardware in form of elementary gates (\ie, transpilation), and
	\item the hardware-related noise of \ac{NISQ} devices as a result of, \eg, decoherence and dissipation effects \cite{burnett2019} as well as \ac{SPAM} errors \cite{mingyu2018}.
\end{itemize}
Consequently, in addition to randomness that occurs in classical \ac{ML} algorithms (\eg, dropout \cite{srivastava2014}), additional randomness may arise in the context of \ac{QML}.
Handling value functions under randomness, noise or uncertainty is therefore essential for deriving meaningful \acp{SV} for \ac{QML}.
Recently, a theoretical framework for such value functions has been proposed \cite{heese2022b}.
Following this reference, we consider an uncertain value function of the form
\begin{equation} \label{eqn:v}
	\RV{v} (\subS) := v (\subS) + \RV{\nu}(\subS)
\end{equation}
based on $v(\subS)$ from \cref{eqn:v:regular} with an additive noise given by the random variable $\RV{\nu}(\subS) \sim \mathbb{Q}(\cdot~|~\subS)$ that follows an arbitrary but fixed probability density $q(\cdot~|~\subS)$ for all $\subS \subseteq \allS$.
Here, $\RV{\nu}(\subS)$ is assumed to cover all sorts of noise which arise from QPU calculations.
This form of additive noise implies a specific structure of the uncertainty under consideration. However, since the additive noise term in \cref{eqn:v} explicitly depends on the coalition $S$, we can in fact use it to formally describe any kind of noisy behavior. As also already pointed out in \cite{heese2022b}, it remains an open research question to evaluate other types of noise representations. In the following, we limit ourselves to the additive form.
{SV}s for such uncertain value functions can then be defined in analogy to \cref{eqn:phi} as an expectation value
\begin{equation} \label{eqn:phi}
	\Phi_i := \expect{ \RV{U}_i }
\end{equation}
of the random variable $\RV{U}_i \sim \mathbb{P}_i$ with probability density
\begin{equation} \label{eqn:deltav:p}
	p_i(u) = \sum_{\subS \subseteq \allS \setminus \{i\}} w(\subS) 
	\int_{\mathbb{R}} q(\nu~|~\subS \cup \{i\}) \ q(\nu + \Delta_i v(\subS) - 
	u~|~\subS) \, \mathrm d \nu,
\end{equation}
where we recall \cref{eqn:w,eqn:deltav}.
It can be shown that \acp{SV} of uncertain value functions as defined in \cref{eqn:phi} represent \emph{regular} \acp{SV} in the sense of \cref{eqn:phi:regular} with a deterministic but shifted value function.
Therefore, all properties of regular \acp{SV} also apply for \acp{SV} of uncertain value functions.
We refer to \cite{heese2022b} and references therein for a more detailed discussion.\par

In practice, the probability density $q(\cdot~|~\subS)$ is typically unknown (\eg, because the statistics of the noise are unknown) and $\Phi_i$ can 
consequently only be estimated.
In the following, we briefly present two possible estimators, which are also outlined in \cite{heese2022b}.
For the first approach, the value function is sampled $K\in\mathbb{N}$ times for every coalition $\subS\in\allS$.
An unbiased estimator for $\Phi_i$ is then given by the sample mean
\begin{equation} \label{eqn:phi:sample}
	\hat{\Phi}_i^{K} := \frac{1}{K} \sum_{\subS \subseteq \allS \setminus 
		\{i\}} w(\subS) \sum_{k=1}^K \left[ \tilde{v}^k (\subS \cup \{i\}) - 
	\tilde{v}^k (\subS) \right],
\end{equation}
where $\tilde{v}^k (\subS)$ represents the $k$th realization of the value function $\RV{v}(\subS) \sim \mathbb{Q}(\ \cdot-v(\subS)\ |\ \subS)$ for $k \in 
\{1,\dots,K\}$.
In total, this approach requires $2^NK$ value function samples.
Due to its linearity, \cref{eqn:phi:sample} can also be understood as a mean of $K$ \acp{SV}, each based on one value function sample.\par

An alternative estimator can be employed without iterating $K$ times over all $2^N$ possible coalitions.
For this purpose, the coalitions themselves are treated as a random variable $\RV{\subS} \sim w$, \cref{eqn:w}, and $n\in\mathbb{N}$ random samples are drawn from the set $\allS \setminus \{i\}$ for each player $i$, which leads to the multiset of realizations $\allS_i^{\mathrm{r}}$.
For each realization $\subS \in \allS_i^{\mathrm{r}}$, the value 
functions $\RV{v}(\subS) \sim \mathbb{Q}(\ \cdot-v(\subS)\ | \ \subS)$ and $\RV{v}(\subS \cup \{i\}) \sim \mathbb{Q}(\ \cdot-v(\subS)\ |\ \subS \cup \{i\})$ are both sampled $K$ times.
In total, $2nK$ value functions are sampled for each player $i$.
Once the sampling for all $N$ players is completed, all value function samples are collected in the multigroup $\mathcal{V}^{\mathrm{r}}$.
Thus, a (possibly empty) multiset of realizations $\mathcal{V}^{\mathrm{r}}(\subS) := \{ \tilde{v}^1 (\subS), \dots, 
\tilde{v}^{\vert\mathcal{V}^{\mathrm{r}}\vert} (\subS) \}$ can be assigned to every coalition $\subS \in \allS$ from the total collection $\mathcal{V}^{\mathrm{r}}$.
The latter corresponds to the additive union $\mathcal{V}^{\mathrm{r}} = \biguplus_{\subS \in \allS} \mathcal{V}^{\mathrm{r}}(\subS)$.
Similar to \cref{eqn:phi:sample}, $\tilde{v}^k (\subS)$ represents the $k$th realization of the value function for $k \in \{1,\dots,\vert\mathcal{V}^{\mathrm{r}}(\subS)\vert\}$, where $\vert\mathcal{V}^{\mathrm{r}}(\subS)\vert \geq 0$ denotes the number of value function samples for the coalition $\subS \in \mathcal{V}^{\mathrm{r}}$.
In conclusion, the expression
\begin{equation} \label{eqn:phi:multisample}
	\hat{\Phi}_i^{n,K} := \frac{1}{n} \sum_{\subS \in 
		\allS_i^{\mathrm{r}}} 
	\left[ \frac{1}{\vert\mathcal{V}^{\mathrm{r}}(\subS \cup \{i\})\vert} 
	\sum_{v 
		\in \mathcal{V}^{\mathrm{r}}(\subS \cup \{i\})} v - 
	\frac{1}{\vert\mathcal{V}^{\mathrm{r}}(\subS)\vert} \sum_{v' \in 
		\mathcal{V}^{\mathrm{r}}(\subS)} v' \right]
\end{equation}
also represents an unbiased estimator for $\Phi_i$.
In total, it requires up to $2nNK$ value function samples.
This estimator can therefore be evaluated with less value function samples than \cref{eqn:phi:sample}, if $n$ is chosen such that $n < 2^{N-1} / N$ (presuming that $K$ is chosen equally).\par
\Acp{SV} are defined as expectation values of random variables.
The corresponding higher moments consequently allow further insight into the probability distribution of these random variables.
For example, the standard deviation
\begin{equation} \label{eqn:phi:V}
	\std{ \Phi_i } := \sqrt{ \expect{ \RV{U}_i^2 } - \Phi_i^2 }
\end{equation}
can be considered as a measure of the effective uncertainty of $\Phi_i$.

\subsection{Shapley values for classical machine learning}
In order to lay out how \acp{SV} can be applied to quantum circuits, we recall their usage in the context of \ac{ML} \cite{rozemberczki2022}. We specifically limit ourselves to two types of applications that are related to our work: (i) \acp{SV} as a model-agnostic \ac{XAI} tool to explain the influence of model inputs on predictions and (ii) \acp{SV} as part of a \ac{NAS} framework with the purpose to analyze the importance of the building blocks of neural networks. We only provide a brief summary of these two applications here, a detailed review of \acp{SV} for \ac{ML} is beyond the scope of this paper.\par
For the first application, the use of \acp{SV} as a model-agnostic \ac{XAI} tool \cite{strumbelj2010,lundberg2017,sundararajan2017,stier2018,xiao2022,oloulade2025}, features (\ie, distinct dimensions of a data point used as the model input) take the role of players and the value function is based on the corresponding model output.
\Acp{SV} of such a coalition game can then be used as an importance measure of each feature. This can be considered as the ``traditional'' usage of \acp{SV} in \ac{XAI}.\par 
Explanations are typically generated for a single data point, consisting of an $N$-dimensional feature vector $\VEC{x} \in \mathbb{R}^N$.
In this case, each feature $i \in \{1,\dots,N\}$ corresponds to a player and the value function is of the form
\begin{equation} \label{eqn:v:xml:single}
	v(\subS) = v(\subS;\VEC{x},f),
\end{equation}
where $f$ denotes the model of interest.
An alternative approach is to aggregate a score over multiple or all data points, which leads to feature importances with respect to an entire data set $\{\VEC{x}_1,\dots,\VEC{x}_d\}$ consisting of data points $\VEC{x}_i \in \mathbb{R}^N$ for $i \in \{1,\dots,d\}$.
Consequently, the value function is of the form
\begin{equation} \label{eqn:v:xml:multi}
	v(\subS) = v(\subS;\{\VEC{x}_1,\dots,\VEC{x}_d\},f)
\end{equation}
in analogy to \cref{eqn:v:xml:single}.\par
In general, value functions of the form presented in \cref{eqn:v:xml:single,eqn:v:xml:multi} can be realized in many different variants leading to \acp{SV} with different interpretations \cite{sundararajan2020}.
To evaluate coalitions of features, a common approach is to eliminate all features not present in a coalition $\subS$, introduced as SHAP by \cite{lundberg2017}.
To this end, the expectation of the model output is taken \wrt all features not present in the coalition, $\mathbb{E}[f(\RV{X})~\vert~x_S]$.
Since computational complexity is a major issue for practical applications \cite{van2022}, this expectation is usually approximated by sampling from a given data set.
Additional assumptions like feature independence and linearity can reduce the computational cost further.
Specialized methods, \eg, KernelSHAP \cite{lundberg2017}, employ locally linear surrogate models to approximate \acp{SV}.
There are also critical aspects of \acp{SV} in this context as for example studied in \cite{kumar2020,huang2023}.\par
For the second application, the use of \acp{SV} as part of a \ac{NAS} framework \cite{leon2015,stier2018,xiao2022,diazortiz2023,oloulade2025}, building blocks of a neural network constitute the players, and the value function is based on the performance of the pruned network, which contains only the building blocks from the respective coalition.\par
Since (game theoretic) \ac{NAS} can be implemented in many different ways \cite{stier2018}, we present here a high-level description that focuses on the key ideas. A typical approach is to choose a specific \ac{ML} task (\eg, image classification on a particular data set) and an appropriate neural network to solve this task. Then, the network is decomposed into $c$ mutually excluding structural components (\eg, layers or parts of layers). Such a decomposition is not unique and each choice leads to a different coalition game. Formally, each structural component $i \in \{1,\dots,c\}$ corresponds to a player and the value function is of the form
\begin{equation} \label{eqn:v:nas}
	v(\subS) = v(\subS;f,t),
\end{equation}
where $f$ denotes the neural network of interest and $t$ the \ac{ML} task (including the  data). To evaluate \cref{eqn:v:nas}, $f$ is first pruned to contain only the structural elements that represent the players in the coalition $\subS$. Then, the pruned model is trained on the task $t$ and a predefined performance metric (\eg, classification accuracy) is determined as the outcome of $v(\subS)$.\par
Within the \ac{NAS} framework, the resulting \acp{SV} can then for example be used to iteratively prune the network by removing the structural components that are associated with the smallest \acp{SV} \cite{leon2015}. As for the first \ac{XAI} application of explaining features, computational complexity is again a major issue for the practical realization of a \ac{SV}-based \ac{NAS}. A common solution is to use a random sampling approximation \cite{fatima2007}.

\subsection{Quantum machine learning} \label{sec:quantum machine learning}
The interdisciplinary field of \ac{QML} lies at the interface of \ac{ML} and quantum computing \cite{biamonte2017,schuld2018}.
It encompasses several research directions that consider data and algorithms that may both be classical, quantum, or a hybrid combination thereof.
In this manuscript, we focus on quantum-enhanced \ac{ML} \cite{dunjko2016}, which considers the use of quantum algorithms to improve upon \ac{ML} tasks for classical data.
This is typically achieved by hybrid algorithms that include a classical and a quantum part.
Such kind of algorithms can already be run on current \ac{NISQ} devices.\par
In {\cref{fig:qml-pipeline}}, we show a simplified sketch of such a \emph{hybrid quantum-classical {\ac{ML}} pipeline}, which we also shortly refer to as \emph{hybrid {\ac{ML}} pipeline} in the following.
Classical data is presented to a classical \emph{host}---typically consisting of (one or more) \acp{CPU}, \acp{GPU}, and \ac{RAM}---, which communicates with a \ac{QPU}.
A quantum circuit is executed on the \ac{QPU}.
This circuit is controlled by features and variational parameters from the classical host.
Features in this context refer to appropriately preprocessed classical data points that can be encoded in the circuit by a suitable combination of gates such that the resulting quantum state contains the respective information, for example in form of amplitudes or angles \cite{weigold2020,schuld2021a}.
We denote a classical $k$-dimensional feature vector by $\VEC{x} \in \mathcal{X} \subseteq \mathbb{R}^k$.\par
\begin{figure}[!t]
	\centering
	\includefigure{1}{./fig\_qml\_pipeline}
	\caption{Simplified sketch of a \emph{hybrid quantum-classical \ac{ML} pipeline} (or \emph{hybrid \ac{ML} pipeline} for short) in form of a variational quantum algorithm.
Preprocessed features $\VEC{x}$ and parameters $\VEC{\theta}$ from a classical host (represented here by a \ac{CPU}) control a variational circuit that is executed on a \ac{QPU}.
The measurement results $\VEC{B}$ are returned to the classical host (in form of bit strings) and enable a quantum classical optimization loop for the parameters.
In the end, a suitable model for the proposed \ac{ML} problem can be realized.
Prospectively, such a pipeline involves multiple \acp{CPU} and \acp{QPU} as well as \acp{GPU} \cite{gambetta2022}.}
	\label{fig:qml-pipeline}
\end{figure}
Furthermore, the variational parameters represent the degrees of freedom of the circuit depending on the chosen ansatz.
An ansatz in this sense describes a particular sequence of gates, where each gate can depend on one or more variational parameters.
We denote a $p$-dimensional vector of variational parameters by $\VEC{\theta} \in \mathcal{P} \subseteq \mathbb{R}^p$.
Quantum circuits depending on such parameters are also known as parameterized or \emph{variational circuits} and are fully defined by a unitary operator $U(\VEC{x}, \VEC{\theta})$ acting on the joint Hilbert space of all $q$ qubits that have initially been prepared in the ground state $\ket{0}$ as shown in \cref{fig:qml-pipeline}.
Hence, the unitary operator yields a final quantum state
\begin{equation} \label{eqn:psiU}
	\ket{\Psi(\VEC{x}, \VEC{\theta})} := U(\VEC{x}, \VEC{\theta}) \ket{0}
\end{equation}
of the joint system of qubits.\par
After execution, the measurement results are returned to the classical host.
For $s$ shots (or repeated circuit evaluations per execution), the measurement results consist of a multiset of $s$ bit strings
\begin{equation} \label{eqn:B}
	\VEC{B}(\VEC{x}, \VEC{\theta}) \equiv \VEC{B} := \{ \VEC{b}_1, \dots, \VEC{b}_s \},
\end{equation}
each of which represents the measured states of all $q$ qubits in form of a vector $\VEC{b}_i \in \{0,1\}^{q} \,\forall\, i \in \{1,\dots,s\}$.
In this sense, each bit string $\VEC{b}_i$ can be considered as a realization of a random variable $\RVEC{b}$ with probability mass
\begin{equation} \label{eqn:m}
	\mathbb{P}_{\VEC{x}, \VEC{\theta}}(\VEC{b}) := \braket{\Psi(\VEC{x}, \VEC{\theta}) | M(\VEC{b}) | \Psi(\VEC{x}, \VEC{\theta})},
\end{equation}
where $M(\VEC{b})$ represents the \ac{POVM} element of the measurement corresponding to the outcome $\VEC{b}$ \cite{nielsen2011}.
Summarized, the variational quantum circuit can be understood as a mapping
\begin{equation} \label{eqn:xtB}
	\VEC{x}, \VEC{\theta} \mapsto \VEC{B}
\end{equation}
from features $\VEC{x}$ and variational parameters $\VEC{\theta}$ to a realization $\VEC{B}$ of a random variable $\RVEC{B}$.\par
Typically, the variational parameters are optimized with a classical optimizer during training such that a suitable model for the proposed \ac{ML} problem can eventually be realized after a certain number of quantum-classical iterations.
Since the variational circuit is of central importance, this kind of hybrid \ac{ML} pipeline is also referred to as \ac{VQA} \cite{cerezo2021}.
The resulting model involves a classical host and (optionally) a \ac{QPU} for inference.\par
Classical \ac{ML} and \ac{QML} problems can be divided into three different categories:
\begin{itemize}
	\item \emph{Supervised learning.} Given a labeled dataset, the goal is to infer the label for unknown samples (see, \eg, \cite{schuld2021b}).
	\item \emph{Unsupervised learning.} The task is to learn patterns of unlabeled data, for example to perform clustering (see, \eg, \cite{kyriienko2022}).
	\item \emph{Reinforcement learning.} An agent is trained to learn actions that maximize a given reward function (see, \eg, \cite{meyer2022}).
\end{itemize}
All three problem categories can be approached with the aforementioned hybrid \ac{ML} pipeline based on variational circuits.\par
For this purpose, measurement results are mapped onto a result value
\begin{equation} \label{eqn:h}
	h: \VEC{B} \mapsto \VEC y \in \mathcal{Y}
\end{equation}
in some solution space $\mathcal{Y}$.
For classification or clustering, $\mathcal{Y}$ may represent a set of class labels, whereas for regression, it can correspond to (a subset of) $\mathbb{R}^t$ for $t$-dimensional targets.
For reinforcement learning, $\mathcal{Y}$ may be a set of feasible policies for an agent.\par
In each of those \ac{ML} scenarios, there are common loss functions (or objectives) which are used to adjust the variational parameters $\VEC{\theta}$ during training, \eg, through gradient descent or evolutionary optimization.
Typically, a loss function
\begin{equation} \label{eqn:loss}
	\ell: \{ (\VEC x_1, \VEC y_1 ), \dots, (\VEC x_d, \VEC y_d ) \}, \vartheta \mapsto \mathbb{R}
\end{equation}
is aggregated over a training dataset $\{ (\VEC x_1, \VEC y_1 ), \dots, (\VEC x_d, \VEC y_d ) \}$ consisting of $d$ tuples of feature vectors $\VEC x_i\in\mathcal{X}$ and corresponding solutions $\VEC y_i\in\mathcal{Y}$ for all $i \in \{1,\dots,d\}$, where each of the latter is determined via \cref{eqn:h} using a respective measurement to obtain \cref{eqn:B}.
Additionally, $\ell$ may depend on meta data $\vartheta$ such as ground-truth labels in supervised learning, depending on context.
Examples for loss functions include mean squared error, cross entropy loss or log-likelihood.

%% file: method.tex
\section{Shapley values for quantum circuits} \label{sec:method}
In \cref{sec:quantum machine learning}, we have outlined a typical hybrid quantum-classical \ac{ML} pipeline in form of a \ac{VQA} as sketched in \cref{fig:qml-pipeline}.
It relies on the execution of variational quantum circuits, which are defined by a unitary operator $U(\VEC{x}, \VEC{\theta})$ acting on the joint Hilbert space of all $q$ qubits that have initially been prepared in the ground state $\ket{0}$.
Moreover, it depends on the $k$-dimensional features $\VEC{x} \in \mathcal{X} \subseteq \mathbb{R}^k$ as well as the $p$-dimensional variational parameters $\VEC{\theta} \in \mathcal{P} \subseteq \mathbb{R}^p$.\par
Since \acp{SV} enable model-agnostic explainability, they can be applied to \ac{QML} models (\ie, models, which result from a quantum-enhanced \ac{ML} pipeline) in complete analogy to the ``traditional'' \ac{XAI} approach for classical \ac{ML}, where features represent players and the model output determines the value function \cite{power2024,gilfoster2024}.\par
In this manuscript, we go beyond this feature-based application and propose a more subtle use of \acp{SV} that is centered around the architecture of \ac{QML} models.
Specifically, presume that a hybrid \ac{ML} pipeline or its resulting model contains a (possibly parameterized) circuit for quantum evaluations.
Our goal is to find an interpretation or explanation for this crucial component within the pipeline or model.
For this purpose, we consider the gates (or groups of gates) as players (or their respective indices, to be more precise) and a value function that is determined by the measurement results.
This approach allows us to assign a \ac{SV} to every gate of the circuit with respect to the chosen value function and hence assign a corresponding contribution.
Consequently, we are able to evaluate the quality of different ansätze and their building blocks in great detail.
We refer to our approach as \acfp{QSV}\footnote{In an earlier version of the manuscript, we used the term \acp{QSVold}. However, we have changed this term to better reflect the fact that we are applying classical (uncertain) Shapley values to quantum gates and quantum-based value functions without actually extending the concept of Shapley values themselves to a quantum version.}.\par
The premise of \acp{QSV} is conceptionally related to \ac{SV}-based \ac{NAS} in the sense that we consider building blocks of \ac{ML} models as players. However, we present our approach in a way that is customly tailored for quantum computations and yet highly generic. Furthermore, we focus on its use for explainability rather than using it within a \ac{QCAS} framework.\par
\begin{figure}[!t]
	\centering
	\includefigure{1}{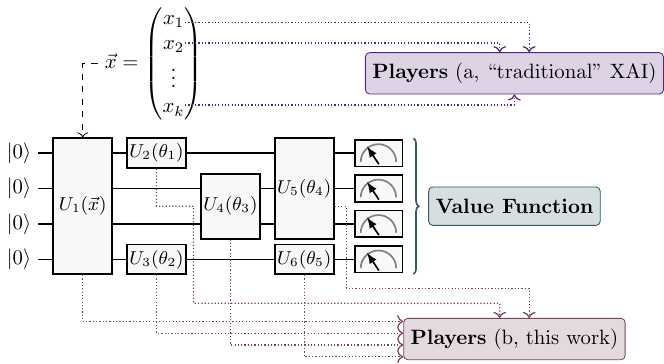}
	\caption{Two possible approaches for \acp{SV} in \ac{QML} involving variational quantum circuits: (a) \acp{SV} from classical \ac{ML} for which features represent players.
(b) Newly proposed \acp{QSV} for which quantum gates represent players.
In both cases, the value function, \cref{eqn:vQ}, is determined by the measurement results.
Shown is an exemplary four-qubit variational circuit consisting of a data encoding unitary for a $k$-dimensional feature vector $\VEC{x} := \{x_1,\dots,x_k\}$ and a set of five parameterized gates as outlined in \cref{fig:qml-pipeline}, where each parameterized gate depends on a single variational parameter $\VEC{\theta}^{i+1} := \{\theta_i\} \subset \VEC{\theta} \,\forall\, i \in \{1,\dots,5\}$ with $\VEC{\theta} := \{\theta_1, \dots, \theta_5\}$.
In terms of \cref{eqn:U}, $G=6$, $U_1(\VEC{x}^1, \VEC{\theta}^1):=U_1(\VEC{x})$, and $U_{1+i}(\VEC{x}^{1+i}, \VEC{\theta}^{1+i}):=U_{1+i}(\theta_i) \,\forall\, i \in \{1,\dots,5\}$.
For both of the \ac{SV} approaches presented here, the gate parameters $\VEC{\theta}$ are assumed to be arbitrary but fixed.}
	\label{fig:shapleys}
\end{figure}

In summary, we identify two approaches for applying \acp{SV} in \ac{QML} as sketched in \cref{fig:shapleys}: either using features as players (the ``traditional'' \ac{XAI} approach) or using gates as players (what we consider in this work with \acp{QSV}).
Both concepts found on the same premise of a coalition game.\par
Game theory in the context of quantum computing has been considered before; see, \eg, \cite{khan2018} for a recent review of (non-cooperative) quantum games.
\Acp{QSV} allow to assign a contribution to individual gates as a whole, where the gates can either be parameterized with a fixed parameter value or non-parameterized.
This is in contrast to other approaches that measure the importance of the gate parameters themselves, for example with gradient-based methods \cite{haug2021}.
\par
In the remaining part of this section, we first introduce a formal definition of \acp{QSV}.
Next, we motivate and present a selection of possible value functions that can be used for various \ac{QML} use cases.
We conclude with a brief summary of the key ingredients of our proposed method.

\subsection{Definition of SVQXs}
To implement \acp{QSV}, we decompose the unitary operator
\begin{equation} \label{eqn:U}
	U(\VEC{x}, \VEC{\theta}) = \prod_{g \in \{1,\dots,G\}} U_g(\VEC{x}^g, \VEC{\theta}^g) := U_G(\VEC{x}^G, \VEC{\theta}^G) \cdots U_1(\VEC{x}^1, \VEC{\theta}^1)
\end{equation}
from the quantum circuit of interest, \cref{eqn:psiU}, into $G$ unitary 
operators $U_g(\VEC{x}^g, \VEC{\theta}^g)$ that represent gates (or groups of 
gates) within the circuit and may depend on (a possibly empty subset of) 
features $\VEC{x}^g \subseteq \VEC{x}$ and variational parameters 
$\VEC{\theta}^g \subseteq \VEC{\theta}$, respectively, for all $g \in 
\{1,\dots,G\}$.
The product in \cref{eqn:U} is to be understood in such a way that the terms are ordered as indicated.\par
Furthermore, we define a set
\begin{equation} \label{eqn:A}
	A := \{A_1,\dots,A_{N}\} \subseteq \{1,\dots,G\}
\end{equation}
of active gates, which participate as players, whereas the remaining gates
\begin{equation} \label{eqn:R}
	R := \{1,\dots,G\} \setminus A
\end{equation}
are treated as passive (or remaining) part of the circuit. 
%are not treated as players.
The decomposition of the unitary operator into gates and the respective subset of active gates can be chosen at will for the use case of interest.\par
\begin{figure}[!t]
	\centering
	\includefigure{\scaleB}{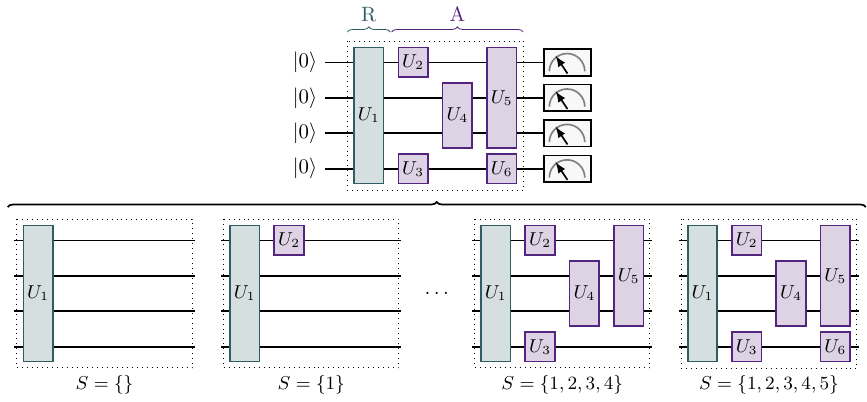}
	\caption{Schematic representation of the set of circuits that are involved in a	coalition game for \acp{QSV}.
Exemplarily, we consider the circuit from \cref{fig:shapleys} as the circuit of interest with an active gate set $A=\{A_1=2,A_2=3,A_3=4,A_4=5,A_5=6\}$, \cref{eqn:A}, and a set of remaining gates $R=\{1\}$, \cref{eqn:R}.
The original circuit of interest is shown on top, it contains the six gates $U_1,\dots,U_6$, where we omit all arguments.
The index of each gate indicates the corresponding gate index $g \in \{1,\dots,G\}$ with $G=6$ in \cref{eqn:U}.
Each coalition $\subS$ can be associated with a circuit that contains the gates 
whose indices are included in the set $\{ A_a \,|\, a \in \subS\} \cup R$ as 
defined in \cref{eqn:USR}.
Four exemplary coalitions and their corresponding circuits are shown at the bottom.
The leftmost and rightmost circuits are associated with the empty and grand coalition, respectively.}
	\label{fig:qshap-circuits}
\end{figure}

The corresponding coalition game involves $N=\vert A \vert$ players.
Coalitions in this game represent sets of active gates that can be associated with circuits consisting of the gates of the coalition and the passive gates in $R$.
For a given coalition $\subS$, such a circuit is defined by the unitary operator
\begin{equation} \label{eqn:USR}
	U(\subS, R, \VEC{x}, \VEC{\theta}) := \prod_{\substack{g \in \{ A_a \,|\, a 
	\in \subS\} \cup R}} U_g(\VEC{x}^g, \VEC{\theta}^g)
\end{equation}
that yields a quantum state
\begin{equation} \label{eqn:psiUSR}
	\ket{\Psi(\subS, R, \VEC{x}, \VEC{\theta})} := U(\subS, R, \VEC{x}, 
	\VEC{\theta}) \ket{0}
\end{equation}
in analogy to \cref{eqn:psiU}.
The product in \cref{eqn:USR} is to be understood as ordered in the same way as in \cref{eqn:U}.
Measurement results for $q$ qubits are consequently given by
\begin{equation} \label{eqn:BSR}
	\VEC{B}(\subS, R, \VEC{x}, \VEC{\theta}) \equiv \VEC{B}(\subS) := \{ 
	\VEC{b}_1(\subS), \dots, \VEC{b}_s(\subS) \},
\end{equation}
in analogy to \cref{eqn:B} with $\VEC{b}_i(\subS) \in \{0,1\}^{q}$ and
\begin{equation} \label{eqn:bSR}
	\VEC{b}_i(\subS) \sim \mathbb{P}_{\subS, R, \VEC{x}, \VEC{\theta}}(\VEC{b}) 
	:= 
	\braket{\Psi(\subS, R, \VEC{x}, \VEC{\theta}) | M(\VEC{b}) | \Psi(\subS, R, 
	\VEC{x}, \VEC{\theta})}
\end{equation}
for $i \in \{1,\dots,s\}$ in analogy to \cref{eqn:m}.
Hence, \cref{eqn:xtB} can be written as
\begin{equation}
	\subS, R, \VEC{x}, \VEC{\theta} \mapsto \VEC{B}(\subS),
\end{equation}
which also depends on the active gates (players) in $\subS$ and the passive 
gates in $R$.
In \cref{fig:qshap-circuits}, we exemplarily outline the set of circuits that are involved in such a coalition game.\par
Technically, \acp{QSV} are computed in the same way as conventional \acp{SV} via \cref{eqn:phi}.
Their premise is a coalitional game based on gates within a quantum circuit as players.
For such a coalition game, the value function, \cref{eqn:v}, takes the form
\begin{equation} \label{eqn:vQ}
	v(\subS) = v(\subS; R, \VEC{x}, \VEC{\theta}, \VEC{B}(\subS)).
\end{equation}
More generally, the value function can in analogy to \cref{eqn:v:xml:multi} 
depend on multiple feature vectors $\VEC{x}_i \in \mathcal{X}$ and variational 
parameters $\VEC{\theta}_i \in \mathcal{P}$.
Measurement results are then 
obtained for each $\VEC{B}(\subS, R, \VEC{x}_i, \VEC{\theta}_i)$, 
\cref{eqn:BSR}, for all $i \in \{1,\dots,d\}$, hence
\begin{equation} \label{eqn:vQ:mult}
	v(\subS) = v(\subS; R, \Xi_d).
\end{equation}
Here, we made use of the abbreviation
\begin{equation} \label{eqn:vQ:mult:Xi}
	\Xi_d := \{(\VEC{x}_1, \VEC{\theta}_1,\VEC{B}(\subS, R, \VEC{x}_1, 
	\VEC{\theta}_1)),\cdots,(\VEC{x}_d, \VEC{\theta}_d,\VEC{B}(\subS, R, 
	\VEC{x}_d, \VEC{\theta}_d))\}.
\end{equation}
For $d=1$, \cref{eqn:vQ:mult} reduces to \cref{eqn:vQ}.
A value function may also depend on additional meta data.\par
In the most general case of value functions, the measurement results, 
\cref{eqn:BSR}, may refer not just to direct measurements of the circuit of 
interest, but instead to an alternation or modification of this circuit in the 
sense that the unitary from \cref{eqn:USR} is mapped onto another unitary, \ie, 
$U(\subS, R, \VEC{x}, \VEC{\theta}) \rightarrow U'(\subS, R, \VEC{x}, 
\VEC{\theta})$.
This allows to realize experiments like a SWAP test \cite{barenco1997,buhrmann2001}.
The results may also depend on a particular order, for example to realize an iterative quantum state tomography algorithm \cite{zambrano2020}.
In the next section, we discuss examples for value functions, which also include such special cases.\par
In addition to executing circuits on a \ac{QPU}, we also consider the case where a circuit is simulated on classical hardware.
Such an approach allows to perform quantum computations that are not yet feasible on \ac{NISQ} devices due to hardware-related imperfections like dissipation and decoherence \cite{zhou2020}.
There are two major differences between simulating a circuit classically and running it on an actual \ac{QPU}.
Firstly, the simulation is noise-free (but noise can optionally be modeled \cite{georgopoulos2021}).
And secondly, the state $\ket{\Psi(\subS, R, \VEC{x}, \VEC{\theta})}$ from 
\cref{eqn:psiUSR} can be determined directly without the need for a measurement.
In terms of the presented formalism, a simulation corresponds to the case of 
infinite shots, \ie, $s \rightarrow \infty$ in \cref{eqn:B} such that the 
probability distribution $\mathbb{P}_{\subS, R, \VEC{x}, \VEC{\theta}}$, 
\cref{eqn:bSR}, can be obtained with arbitrary precision for all bit strings 
$\VEC{b} \in \{0,1\}^{q}$.
Alternatively, a value function for simulated circuits may also directly depend on the quantum state instead of the measurement results.
An exemplary simulation-based value function is provided in the following section as well. 

\subsection{Value functions for SVQXs} \label{sec:value functions for quantum shapley values}
Value functions for \acp{QSV}, \cref{eqn:vQ:mult}, can be chosen in many different ways.
First of all, value functions from classical \ac{XAI} can be used in the same way for \ac{QML} tasks.
In analogy to \cref{eqn:v:xml:single,eqn:v:xml:multi}, we can define
\begin{equation} \label{eqn:v:qml:single}
	v(\subS) = v(\subS;\VEC{x},h(\VEC{B}(\subS, R, \VEC{x}, \VEC{\theta})),
\end{equation}
for single data points $\VEC{x} \in \mathbb{R}^N$ and
\begin{equation} \label{eqn:v:qml:multi}
	v(\subS) = v(\subS;\VEC{x}_1,\dots,\VEC{x}_d,h(\VEC{B}(\subS, R, \VEC{x}_1, 
	\VEC{\theta}_1),\dots,h(\VEC{B}(\subS, R, \VEC{x}_d, \VEC{\theta}_d))
\end{equation}
for a data set $\VEC{x}_1, \dots, \VEC{x}_d$ consisting of $d$ data points $\VEC{x}_i \in \mathbb{R}^N$ for $i \in \{1,\dots,d\}$.
Here, we recalled $h$ from \cref{eqn:h}.
For example, a loss function of the form of \cref{eqn:loss} can be used as the value function in \cref{eqn:v:qml:multi}, that is,
\begin{equation} \label{eqn:v:loss}
	v(\subS) = \ell( \{( \VEC{x}_1, h(\VEC{B}(\subS, R, \VEC{x}_1, 
	\VEC{\theta}_1)) ), \dots, (\VEC{x}_d, h(\VEC{B}(\subS, R, \VEC{x}_d, 
	\VEC{\theta}_d)) ) \}, \vartheta).
\end{equation}
\acp{QSV} based on such a value function allow to assign a contribution of each gate to the loss.\par
Moreover, more specialized value functions can be used independent of a particular \ac{ML} task to evaluate certain properties of a quantum circuit.
In this context, we specifically propose four typical use cases in form of questions:
\begin{enumerate}
	\item How uniformly is the unitary space explored by an ansatz circuit?
	\item What ability has an ansatz circuit to create entangled states?
	\item How does a circuit evaluation perform on \ac{NISQ} hardware in comparison with an idealized classical simulator?
	\item How efficiently can a circuit be executed on a specific \ac{NISQ} hardware device?
\end{enumerate}
In the following, be briefly outline potential value functions to answer these questions.\par
For the first use case, a suitable metric to quantify the uniform exploration capability of the unitary space is given by the \emph{expressibility} \cite{sim2019}.
It corresponds to the distance between the distribution of unitaries generated by the ansatz with randomly chosen variational parameters and the maximally expressive uniform distribution of unitaries.
We consider an ansatz circuit with a corresponding unitary, \cref{eqn:psiU}, that depends on a parameter vector $\VEC{\theta}$, whereas the feature vector $\VEC{x}$ is presumed to be constant.
An estimator for the expressibility of this circuit 
$\eta(\subS,R,\VEC{x},c_p,c_b,\mathcal{F})$ is presented in 
\cref{sec:app:expressibility}.
Here, $c_p$ and $c_b$ denote parameters that determine the precision of the estimator, whereas $\mathcal{F}$ contains measurement results from SWAP tests.
The corresponding value function
\begin{equation} \label{eqn:v:expressibilty}
	v(\subS) = -\eta(\subS,R,\VEC{x},c_p,c_b,\mathcal{F}) \leq 0
\end{equation}
allows to quantify the contribution of each gate to the expressibility, \ie, its exploration capability of the unitary space.
This quantity is in particular independent of a specific parameter vector $\VEC{\theta}$.
The negative sign in \cref{eqn:v:expressibilty} ensures that larger values indicate a higher expressibility, whereas smaller values indicate a lower expressibility.\par
To answer the question of the second use case, the \emph{entangling capability} represents a suitable metric \cite{sim2019}.
As we present in \cref{sec:app:entangling capability}, it is based on the mean Meyer-Wallach entanglement measure \cite{meyer2002} for randomly drawn variational parameters.
The metric can be directly used as a value function
\begin{equation} \label{eqn:v:entangling}
	v(\subS) = \lambda(\subS,R,\VEC{x},c_s,\mathcal{T}) \in [0,1]
\end{equation}
that allows to quantify the contribution of each gate to the overall entangling capability of the circuit, where larger values indicate more entangling capability.
This quantity is in particular independent of a specific parameter vector $\VEC{\theta}$.
Here, $c_s$ determines the number of drawn parameter samples, whereas $\mathcal{T}$ represents a list of measurement results for quantum state tomography.\par
For the third use case, we evaluate the circuit of interest for a given coalition, \cref{eqn:USR}, both on an actual \ac{QPU} as well as with a (noise-free) simulator running on a classical host.
As briefly outlined in \cref{sec:app:hellinger fidelity}, this approach allows us to use the \emph{Hellinger fidelity} \cite{hellinger1909} to quantify the degree of agreement between the measurement results from the quantum device with the results from the classical simulator based on the respective distributions of bit strings.
Two effects are responsible for a potential deviation between these two results.
First, the limited number of shots on the \ac{QPU} and second, hardware-related uncertainty from the \ac{NISQ} device that is not present on the idealized simulator.
Since a lower degree of agreement consequently results from deficiencies (or inadequacies) of the \ac{QPU}, the Hellinger fidelity is a measure of the severity of these deficiencies.
The Hellinger fidelity can be directly used as a value function
\begin{equation} \label{eqn:v:H}
	v(\subS) = H(\subS, R, \VEC{x}, \VEC{\theta}, \VEC{B}(\subS)) \in [0,1]
\end{equation}
that allows to quantify the contribution of each gate to the quantum hardware deficiencies.
By definition, smaller values indicate larger deficiencies.
Here, $\VEC{B}(\subS)$ denotes the acquired measurement 
results, \cref{eqn:BSR}.
In particular, the evaluation of this value function requires both a simulation of the circuit as well as its execution on a \ac{QPU}.\par
The fourth use case considers the practical challenge of efficiently executing an arbitrary quantum circuit on a \ac{NISQ} device.
As explained in \cref{sec:app:estimated execution efficiency}, any quantum circuit must be subjected to a suitable equivalence transformation to allow execution on a specific hardware device, a process also known as \emph{transpilation} \cite{qiskit2021}.
The transpilation procedure transforms a circuit into a form, where it only contains gates from the set $\mathcal{U}$ of available gates as specified by the hardware device of interest.
In this sense, the transpilation can be understood as a necessary \emph{classical} preprocessing procedure for all \ac{QPU} evaluations.
However, such a procedure is ambiguous by definition and can therefore lead to different gate decompositions with a varying number of gates.
In general, a lower number of gates is preferable since each gate contributes to the noise of the corresponding \ac{QPU} computation.
As briefly outlined in \cref{sec:app:estimated execution efficiency}, the 
estimated execution efficiency $\tau(\subS, R, \mathcal{U}, s_1, s_2)$ 
represents a metric that measures how efficiently a circuit can be executed on 
a specific hardware device by assigning a negative penalty value to each gate 
based on the most favorable transpilation of the circuit.
The penalty value is determined by the penalty parameters $s_1<0$ and $s_2<s_1$ for one-qubit and two-qubit gates, respectively.
The corresponding value function
\begin{equation} \label{eqn:v:transpilation}
	v(\subS) = \tau(\subS, R, \mathcal{U}, s_1, s_2) \leq 0
\end{equation}
can therefore be used to quantify the contribution of each gate to the estimated efficiency of executing the circuit on a specific hardware device, where larger values indicate a higher estimated efficiency.\par
In contrast to use case three, only the efficiency of the transpilation process and not the actual hardware errors that arise during execution are considered.
Therefore, the premise of the final use case is different in comparison with the other three: no \ac{QPU} is used to evaluate the value function, instead the transpilation is run on a \ac{CPU} to determine the estimated efficiency based on the properties of a specific hardware device based on the chosen penalty function.

\subsection{Summary}
In conclusion, \acp{QSV} can be defined in analogy to conventional \acp{SV} via \cref{eqn:phi} based on a coalition game in which certain (parameterized or non-parameterized) gates of a quantum circuit constitute the players, \cref{eqn:A}.
The respective value functions, \cref{eqn:vQ}, involve the unitary operators each coalition, \cref{eqn:USR}, with possible use of results from a \ac{QPU} or a classical host or both.
For the sake of convenience, we use the notation
\begin{equation} \label{eqn:qphi}
	\Phi_{(g)} := \Phi_{i} \,\text{with}\, A_i = g
\end{equation}
to refer to the \ac{QSV} of an active gate with gate index $g$, \cref{eqn:U}, where $A_i \in A$, \cref{eqn:A}.
Examples for value functions based on typical use cases are presented in \cref{eqn:v:loss,eqn:v:expressibilty,eqn:v:entangling,eqn:v:H,eqn:v:transpilation}.
Value functions based on \ac{QPU} calculations may suffer from intrinsic noise and are consequently described by \acp{SV} of uncertain value functions, \cref{eqn:phi}, which can be evaluated, \eg, via \cref{eqn:phi:sample} or \cref{eqn:phi:multisample}.

%% file: experiments.tex
\section{Experiments} \label{sec:experiments}
In this section, we demonstrate the use of \acp{QSV} to evaluate (variational) circuits.
For this purpose, we perform multiple experiments on classical simulators and actual IBM quantum hardware.
In total, we consider five use cases for which we recall a selection of the proposed value functions from \cref{sec:value functions for quantum shapley values}.
Three use cases are from the \ac{QML} domain as a first step towards \ac{XQML}: two classifiers in \cref{sec:quantum support vector machine,sec:qnn}, respectively, and a generative model in \cref{sec:qgan}.
Finally, the last two use cases consider circuit transpilation as a classical preprocessing task in \cref{sec:classical preprocessing} and solving a combinatorial optimization problem with a \ac{VQE} in \cref{sec:qaoa}, which allows us to demonstrate the usability of \acp{QSV} beyond \ac{XQML}.
The goal of this section is to present different application areas for \acp{QSV} without going too deep into possible interpretations or even conclusive explanations of the results.\par 

The experiments have been realized with Python using Qiskit \cite{qiskit2021} and a \ac{QSV} toolbox \cite{heese2023git}.
All data used in the experiments is publicly available online \cite{qxmldata2023,aha2021}.\par
We use a selection of (simulated and \ac{NISQ}) \acp{QPU} to perform quantum computations.
Specifically, as simulators we use (i) an idealized statevector simulator (\QCstatevector), which allows a deterministic evaluation of all amplitudes of the final state and (ii) an idealized shot simulator (\QCaer), which samples a finite number of measurements (shots) from the amplitudes of the final state.
Both simulators are implemented in Qiskit and run on a classical host.
In contrast to the \QCstatevector simulator, the \QCaer simulator has a non-deterministic behavior due to the shot noise.
As \ac{NISQ} devices, we use the \QCehningen and \QCoslo \acp{QPU}.
Both of these \acp{QPU} are superconducting quantum hardware devices provided by IBM Quantum through cloud services \cite{ibmq2022a,ibmq2022b}.
Detailed specifications can be found in \cref{sec:app:backends}.\par 

For the estimation of \acp{QSV}, we use either \cref{eqn:phi:sample} or \cref{eqn:phi:multisample}.
The first approach depends on the chosen number of repetitions $K$ and the second on both $K$ and the number of samples $n$.
For practical purposes, we specify $K$ and a sampling fraction $\alpha \in (0,1]$ to fully define our estimation method.
A choice of $\alpha < 1$ indicates that we use \cref{eqn:phi:multisample} with a number of samples
\begin{equation} \label{eqn:alpha}
	n \equiv n(\alpha) := \lceil\alpha2^{N-1}\rceil,
\end{equation}
whereas $\alpha=1$ indicates that we use \cref{eqn:phi:sample}.

\subsection{Quantum support vector machine} \label{sec:quantum support vector machine}
\begin{figure}[!t]
	\centering
	\includefigure{1}{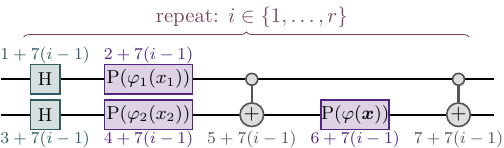}
	\caption{Feature map circuit for the \ac{QSVM}: two-qubit second-order Pauli-Z evolution circuit with $r$ repetitions parameterized by feature vectors $\VEC{x} \in \mathbb{R}^2$.
	We use the circuit symbols from \cref{sec:app:circuit symbols} and the abbreviations from \cref{eqn:phixi,eqn:phix}.
	The number near each gate represents its gate index $g$, \cref{eqn:U}.
	Here and in the following, we omit initialization and measurements in circuit sketches to simplify our presentation.
These operations are by default always realized as shown in \cref{fig:qshap-circuits}.
	}
	\label{fig:kernel-svm:circuit}
\end{figure}

As our first use case, we consider a \ac{QSVM} \cite{havlicek2019} with the task of solving a supervised classification problem on a toy data set with two-dimensional features.
This \ac{QML} model is based on a classical \ac{SVM} \cite{cortes1995} that operates on a kernel matrix from a \ac{QPU}.
That is, the elements of the kernel matrix are estimated from the transition amplitude by measuring the number of all-zero strings of a composition of two consecutive feature map circuits, each parameterized by one of the two feature vectors of interest \cite{havlicek2019}.
Consequently, both training the QVSM and making predictions with it requires the evaluation of quantum circuits.
As feature map circuit, we use a two-qubit second-order Pauli-Z evolution circuit with $7r$ gates, where $r$ denotes the number of repetitions as shown in \cref{fig:kernel-svm:circuit}.
We particularly consider $r \in \{1,2,3\}$.
For the parameterization of the circuit (with two-dimensional feature vectors $\VEC{x} \in \mathbb{R}^2$), we use the abbreviations
\begin{equation} \label{eqn:phixi}
	\varphi_i(x_i) := 2 x_i
\end{equation}
for $i \in \{1,2\}$ and
\begin{equation} \label{eqn:phix}
	\varphi(\VEC{x}) := 2 (\pi - x_1) (\pi - x_2),
\end{equation}
respectively.
For a recent review of QSVMs, we refer to \cite{gentinetta2022} and references therein.\par
\begin{figure}[!t]
	\centering
	\includefigure{1}{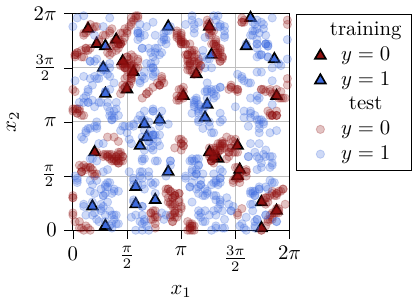}
	\caption{Artificial test and training data sets for the supervised classification problem with two-dimensional features $\VEC{x} \in (0,2\pi]^2$ and binary labels $y \in \{0,1\}$.
We use a \ac{QSVM} to solve this problem.}
	\label{fig:kernel-svm:data}
\end{figure}

For training and testing purposes, we use the artificial data set from \cite{havlicek2019} with two-dimensional features $\VEC{x} \in (0,2\pi]^2 \subset \mathbb{R}^2$ and binary class labels $y \in \{0,1\}$.
By construction, this data can in principle be fully separated by the chosen feature map with $r=2$.
We choose a separation gap of \num{0.3}, which is a hyperparameter that determines the geometrical shape of the data.
In total, we uniformly draw \num{40} training points and \num{1000} test points from the feature space, both with an equal distribution of zero and one class labels, as shown in \cref{fig:kernel-svm:data}.\par
For this use case, all gates are active gates such that $A=\allS$ and 
$R=\varnothing$ as defined in \cref{eqn:A,eqn:R}.
Furthermore, $\Phi_{(g)} = 
\Phi_{i}$ according to \cref{eqn:qphi}.
The considered value function 
$v(\subS)$ is defined by the following procedure:
\begin{enumerate}
	\item Evaluate the kernel matrix for the training data set based on the 
	feature map circuit, \cref{fig:kernel-svm:circuit}, that contains only 
	gates from $\subS$, \cref{eqn:USR}.
	\item Train the \ac{QSVM} using this kernel matrix.
The training algorithm is deterministic.
	\item Evaluate the accuracy of the trained model on the test data set to 
	determine $v(\subS) = \mathrm{acc}$ using the same feature map circuit as 
	before.
\end{enumerate}
The accuracy is defined in the usual sense as
\begin{equation} \label{eqn:acc}
	\mathrm{acc} := \frac{\text{number of correctly predicted test labels}}{\text{total number of test points}} \in [0,1]
\end{equation}
and is therefore of the form of \cref{eqn:v:loss}.
This value function allows to quantify the contribution of each gate to the kernel matrix for a successful training of the \ac{QSVM}, where larger values indicate a better model performance.
\par
\begin{figure}[!t]
	\centering
	\includefigure{\scaleA}{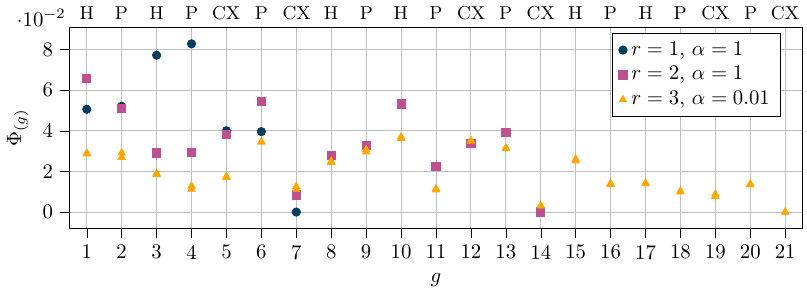}
	\caption{\Acp{QSV} $\Phi_{(g)}$, \cref{eqn:qphi}, for the feature map circuit of a \ac{QSVM}, \cref{fig:kernel-svm:circuit}, with a value function corresponding to the retrained test accuracy of the corresponding model.
A larger \ac{QSV} is consequently more favorable.
We consider three different numbers of repetitions $r \in \{1,2,3\}$.
	The gate index $g$, \cref{eqn:U}, is shown on the bottom, whereas the corresponding gate name is shown on the top of the plot.
	All evaluations are run on the \QCstatevector simulator, which leads to deterministic value functions.
We therefore use $K=1$ repetitions for their evaluation. 
	For $r \in \{1,2\}$, we decide not to sample value functions by choosing a sample fraction $\alpha=1$, \cref{eqn:alpha}, which leads to deterministic \acp{QSV} via \cref{eqn:phi:sample}.
For $r=3$, we choose a value function sample with $\alpha=\num{0.01}$ and perform two independent runs, for each of which we draw different samples using \cref{eqn:phi:multisample}.
We plot the results of both runs with the same symbols.}
	\label{fig:kernel-svm:results:phi}
\end{figure}

We run all evaluations on the \QCstatevector simulator.
Since the training algorithm of the \ac{QSVM} is also deterministic, no randomness is involved in the evaluation of the value function with the consequence that $\hat{\Phi}_i^{1} = \Phi_i = \phi_i$ according to \cref{eqn:phi:sample,eqn:phi,eqn:phi:regular}.
For $r \in \{1,2\}$, we therefore use $K=1$ and $\alpha=1$, \cref{eqn:alpha}.
On the other hand, we choose $K=1$ and $\alpha=\num{0.01}$ for $r = 3$ to reduce the computational effort, \cref{eqn:phi:multisample}, and perform two independent runs, for each of which we draw different samples.
The resulting \acp{QSV} for the feature map circuit, \cref{fig:kernel-svm:circuit}, with different numbers of repetitions $r$ are presented in \cref{fig:kernel-svm:results:phi}.
The resulting test accuracies read $\mathrm{acc}\approx\num{0.842}$ for $r=1$, $\mathrm{acc}\approx\num{0.986}$ for $r=2$, and $\mathrm{acc}\approx\num{0.913}$ for $r=3$.\par
We find that the resulting \acp{QSV} exhibit a rich structure.
First of all, we can observe that the \acp{QSV} for the last $\CNOT$ gate of a feature map circuit vanish since it has no effect on the measured number of all-zero strings and therefore on the value function.
Intermediate $\CNOT$ gates, on the other hand, have non-vanishing \acp{QSV}.
Overall, the $\CNOT$ gates at the end of a repetition (\ie, $g \in \{7,14,21\}$) have small \acp{QSV}.
Supplementary, we show the corresponding distributions of marginal contributions for the evaluation of the \acp{QSV} from \cref{fig:kernel-svm:results:phi} in \cref{sec:app:qsvm marginal contributions}.\par
\begin{figure}[!t]
	\begin{subfigure}[t]{.29\linewidth}
		\centering
		\includefigure{\scaleB}{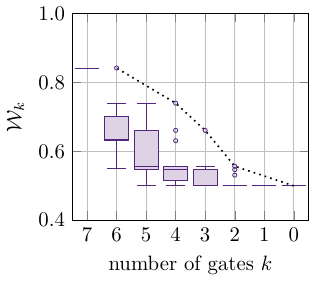}
		\caption{$r=1$, $\alpha=1$}
		\label{fig:kernel-svm:results:mem:1}
	\end{subfigure}%
	\hfill
	\begin{subfigure}[t]{.69\linewidth}
		\centering
		\includefigure{\scaleB}{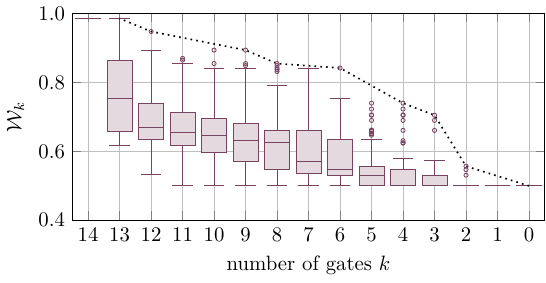}
		\caption{$r=2$, $\alpha=1$}
		\label{fig:kernel-svm:results:mem:2}
	\end{subfigure}%
	\\[.25cm]
	\begin{subfigure}[t]{1.\linewidth}
		\centering
		\includefigure{1}{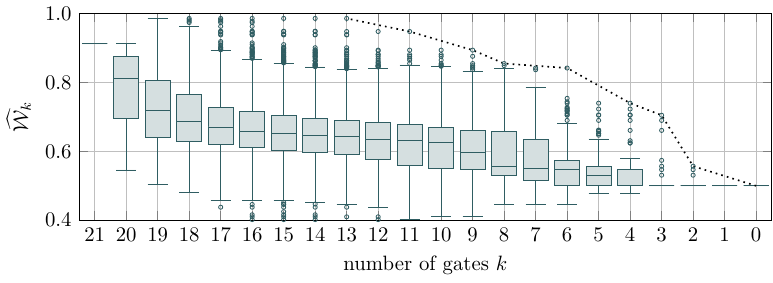}
		\caption{$r=3$, $\alpha=\num{0.01}$}
		\label{fig:kernel-svm:results:mem:3}
	\end{subfigure}%
	\caption{Distribution of value functions $\mathcal{W}_k$, \cref{eqn:W}, for different number of gates $k$ in form of box plots.
The results are obtained for the experiments from \cref{fig:kernel-svm:results:phi}.
We show three different numbers of repetitions $r \in {1,2,3}$.
All evaluations are run on a \QCstatevector simulator with $K=1$.
The dotted line indicates the Pareto frontier for a multi-criteria circuit pruning problem with the goal to minimize the number of gates and to maximize the test accuracy of the \ac{QSVM}.
For $r=3$, we consider the distribution of sampled value functions $\widehat{\mathcal{W}}_k$ for one run.
The corresponding Pareto frontier is an approximation since not all value functions have been evaluated.}
	\label{fig:kernel-svm:results:mem}
\end{figure}

Another perspective on \acp{QSV} is to analyze the corresponding value function distributions for subsets of gates.
For this purpose, we consider the multiset of value functions
\begin{equation} \label{eqn:W}
	\mathcal{W}_k := \{ v \,|\, v = v(\subS) \in \mathbb{R} \,\forall\, \subS 
	\subseteq \allS \land \vert \subS \vert = k \}
\end{equation}
given $k$ (active) gates.
That is, every element in $\mathcal{W}_k$ corresponds to a retrained test accuracy for a \ac{QSVM} with a specific feature map circuit (consisting of $k$ gates in total).
To determine $\mathcal{W}_k$, all value functions have to be evaluated, which corresponds to $\alpha=1$.
In case of $\alpha<1$, we consider the multiset of sampled value functions $\widehat{\mathcal{W}}_k$ instead.
We plot the distribution of $\mathcal{W}_k$ and $\widehat{\mathcal{W}}_k$, respectively, for the results from \cref{fig:kernel-svm:results:phi} as box plots in \cref{fig:kernel-svm:results:mem}.\par
The plots confirm that one gate (the last $\CNOT$) can be removed without lowering the test accuracy.
Furthermore, for $r=3$ removing up to eight gates can even lead to an improvement of the score.
The trade-off between the number of gates and the test accuracy can be considered as a multi-criteria optimization problem for circuit pruning \cite{sim2021}.
The corresponding Pareto frontier is indicated as a dotted line in \cref{fig:kernel-svm:results:mem}.
This analysis can be understood as a very simple form of \ac{QCAS} that allows to select the optimal gate configuration for a certain task.\par
So far, \acp{QSV} have allowed us to gain certain insights into the influence of the gates of a feature map circuit on the test accuracy of a corresponding \ac{QSVM}.
In order to investigate the significance of these insights and the general robustness of \acp{QSV}, we conduct three additional tests in the following.\par
\begin{figure}[!t]
	\centering
	\begin{subfigure}[t]{1.\linewidth}
		\centering
		\includefigure{\scaleC}{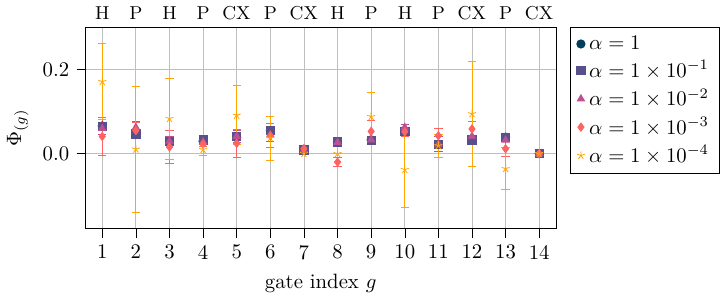}
		\caption{$r=2$}
		\label{fig:kernel-svm:results:phifrac:2}
	\end{subfigure}%
	\\[.25cm]
	\begin{subfigure}[t]{1.\linewidth}
		\centering
		\includefigure{\scaleC}{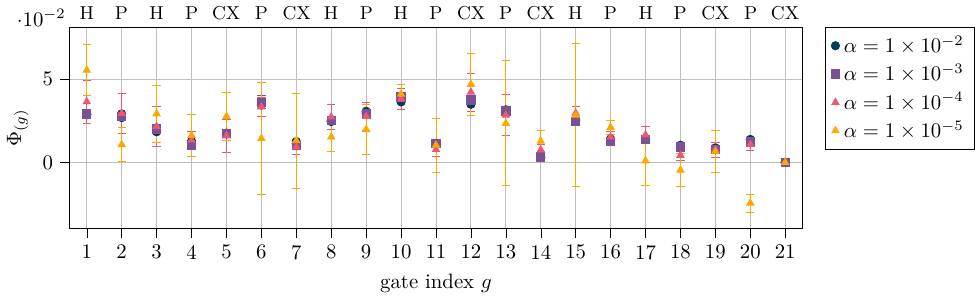}
		\caption{$r=3$}
		\label{fig:kernel-svm:results:phifrac:3}
	\end{subfigure}%
	\caption{Influence of the number of sampled value functions that is determined by the sampling fraction $\alpha$, \cref{eqn:alpha}, on the \acp{QSV} from \cref{fig:kernel-svm:results:phi} for two repetitions $r \in \{2,3\}$.
For $r=2$ and $\alpha=1$, no randomness is involved so that we can show the exact results.
For $r=3$ and $\alpha=\num[scientific-notation=true]{0.01}$, we plot both results of two independent runs for each case with the same symbols.
In all other cases, we show the mean and one standard deviation over three independent runs.
The evaluations are performed on a \QCstatevector simulator with $K=1$.}
	\label{fig:kernel-svm:results:phifrac}
\end{figure}

 
First, we determine the influence of the sampling fraction $\alpha$, \cref{eqn:alpha}, that determines the number of sampled value functions.
For this purpose, we perform three evaluations of the \acp{QSV} from \cref{fig:kernel-svm:results:phi} using different values of $\alpha$ and $K=\num{1}$ for each of the repetitions $r \in \{2,3\}$.
Since the evaluations are performed on a \QCstatevector simulator, the only source of randomness originates from the sampling of value functions.
The results are presented in \cref{fig:kernel-svm:results:phifrac}.
\par
We find that the mean values of the \acp{QSV} converge for increased values of $\alpha$, whereas their standard deviation decreases, as expected.
The standard deviation is in particular different for each gate.\par
\begin{figure}[!t]
	\centering
	\includefigure{1}{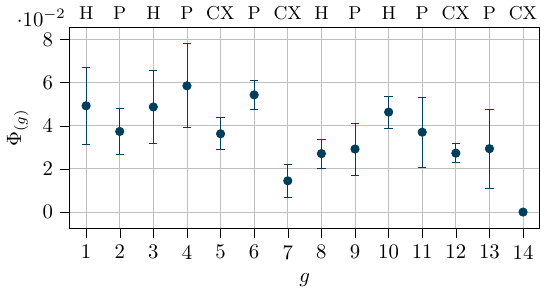}
	\caption{Influence of the data sampling on the resulting \acp{QSV}.
We plot \acp{QSV} for the \ac{QSVM} with $r=2$ in analogy to \cref{fig:kernel-svm:results:phi} based on five different test and training data sets (obtained from sampling the data distribution with different random seeds).
We show the mean and one standard deviation.
The evaluations are performed on a \QCstatevector simulator with $K=1$ and $\alpha=1$.}
	\label{fig:kernel-svm:results:phi-data}
\end{figure}

 
As a second experiment, we test the generality and robustness of the result by evaluating the \acp{QSV} from \cref{fig:kernel-svm:results:phi} for different data sets.
Specifically, we consider the previously used training and test data sets and, in addition, four other training and test data sets of the same size that have been sampled from the same data distribution with different random seeds.
We perform the evaluations on a \QCstatevector simulator with $K=1$ and $\alpha=1$, which eliminates all randomness.
The results are presented in \cref{fig:kernel-svm:results:phi-data}.\par 
From these results it becomes apparent that the \acp{QSV} for some gates are more susceptible to a change of the data than others.
In particular, for $g=4$, $g=11$, and $g=13$ we find a large standard deviation.
This is no surprise since all of these gates are $\PHASE$ gates that are directly influenced by the choice of data.
Moreover, the initial $\HADAMARD$ gates with $g=1$ and $g=3$ show a similarly large standard deviation, indicating that their contribution is also data-dependent.\par
\begin{figure}[!t]
	\centering
	\includefigure{1}{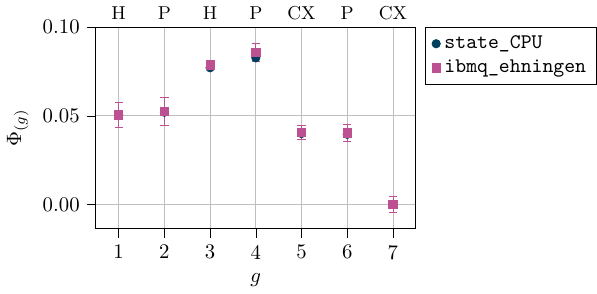}
	\caption{Influence of uncertainty from \ac{NISQ} hardware on the resulting \acp{QSV}.
We plot \acp{QSV} for the \ac{QSVM} with $r=1$ in analogy to \cref{fig:kernel-svm:results:phi}.
All results are evaluated either on a \QCstatevector simulator or the \QCehningen \ac{QPU} with $K=1$ and $\alpha=1$.
For \QCehningen, we show the mean value and five standard deviations over six independent runs.}
	\label{fig:kernel-svm:results:phi-backend}
\end{figure}

Finally, in a third experiment we investigate the influence of noise from a \ac{NISQ} \ac{QPU}.
To this end, we choose $r=1$ and perform six evaluations on the \QCehningen \ac{QPU} with $K=1$ and $\alpha=1$.
Therefore, the only source of randomness are errors from the \ac{QPU}.
These errors (unavoidably) lead to uncertain value functions in the sense of \cref{eqn:v}.
For comparison, we also perform an evaluation on the \QCstatevector simulator.
The results are presented in \cref{fig:kernel-svm:results:phi-backend}.\par
Clearly, the results show that the influence of the errors from the \ac{NISQ} \ac{QPU} on the \acp{QSV} is not significant.

\subsection{Quantum neural network} \label{sec:qnn}
\begin{figure}[!t]
	\centering
	\includefigure{1}{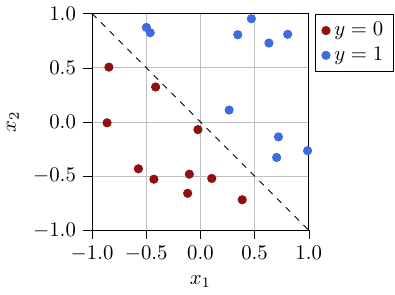}
	\caption{Toy data for the supervised classification problem consisting of \num{20} data points with two-dimensional features $\VEC{x} \in [-1,1]^2$ and binary labels $y \in \{0,1\}$.
We use a \ac{QNN} to solve this problem.}	
	\label{fig:qnn:data}
\end{figure}

\begin{figure}[!t]
	\centering
	\includefigure{1}{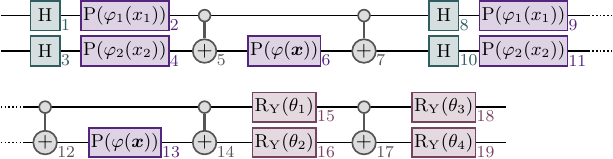}
	\caption{Circuit for the \ac{QNN} parameterized by trainable parameters $\VEC{\theta} \in \mathbb{R}^4$ and feature vectors $\VEC{x} \in \mathbb{R}^2$.
	We use the circuit symbols from \cref{sec:app:circuit symbols} and the abbreviations from \cref{eqn:phixi,eqn:phix}.
	The number near each gate represents its gate index $g$, \cref{eqn:U}.}
	\label{fig:qnn:circuit}
\end{figure}

As our second use case, we consider a \ac{QNN} with the task to perform a binary classification of a toy data set with two-dimensional features.
The data set contains \num{20} data points and is visualized in \cref{fig:qnn:data}.
By construction, it is linearly separable in the feature space.
For reasons of simplicity, we use the same data set for test and training purposes.
The chosen quantum circuit of the \ac{QNN} is shown in \cref{fig:qnn:circuit}.
It is parameterized both by feature vectors $\VEC{x} \in \mathbb{R}^2$ and trainable parameters $\VEC{\theta} \in \mathbb{R}^4$.
The prediction of the class label is determined by the binary measurement result of the first qubit.\par
We train the \ac{QNN} once on the proposed data set with a COBYLA optimizer \cite{powell2007} to determine a fixed value for the parameters $\VEC{\theta} = \VEC{\theta}_{\mathrm{trained}} \approx (\num{3.860}, \num{-1.070}, \num{-1.583}, \num{0.860})$.
The resulting accuracy of the classifier, \cref{eqn:acc}, is $80\%$.
For the evaluation of \acp{QSV}, we choose the active gate set $A =\{2,4,5,\dots,19\}$, \cref{eqn:A}, and a set of remaining gates $R=\{1,3\}$, \cref{eqn:R}.
Hence, all gates except for the initial $\HADAMARD$ gates are treated as active.
The considered value function $v(\subS)$ is defined by the one-shot test 
accuracy of the \ac{QNN} circuit that contains only gates from $S \cup R$, 
\cref{eqn:USR}. 
That is, we perform one shot for every data point to predict the class label (by parameterizing the \ac{QNN} circuit with the current feature vector as $\VEC{x}$ and the trained parameters $\VEC{\theta}_{\mathrm{trained}}$) and evaluate the accuracy via \cref{eqn:acc}.
Using the one-shot test accuracy allows us to investigate the shot noise further below.\par
\begin{figure}[!t]
	\centering
	\includefigure{1}{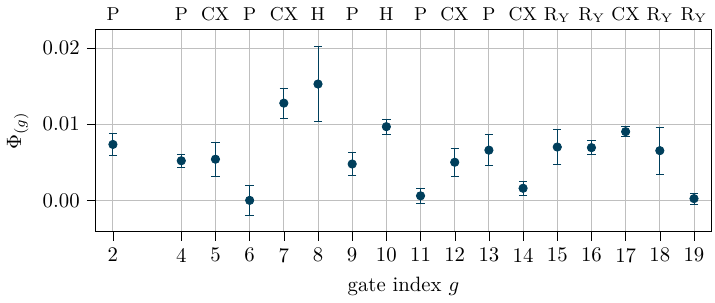}
	\caption{\Acp{QSV} $\Phi_{(g)}$, \cref{eqn:qphi}, for the feature map circuit of a \ac{QNN}, \cref{fig:qnn:circuit}, with a value function corresponding to the one-shot accuracy of the \ac{QNN} without retraining.
A larger \ac{QSV} is consequently more favorable.
In analogy to \cref{fig:kernel-svm:results:phi}, the gate index $g$, \cref{eqn:U}, is shown on the bottom, whereas the corresponding gate name is shown on the top of the plot.
For the gates with $g \in R=\{1,3\}$, no \ac{QSV} is determined.
All evaluations are run on the \QCaer simulator, which includes (simulated) shot noise.
We use $K=32$, $\alpha = \num{0.01}$, \cref{eqn:alpha}, and show mean and one standard deviation over five independent runs.}
	\label{fig:qnn:results:phi}
\end{figure}

We run all evaluations on the \QCaer simulator.
Therefore, the only source of randomness is the (simulated) shot noise.
The shot noise leads to uncertain value functions in the sense of \cref{eqn:v}.
The resulting \acp{QSV} obtained for $K=32$ and $\alpha = \num{0.01}$, \cref{eqn:alpha}, are presented in \cref{fig:qnn:results:phi}.\par
We find that \acp{QSV} show significant differences between certain gates.
In particular, the gates with $g\in\{7,8\}$ attain particularly large \acp{QSV}, whereas the gates with $g \in \{6,11,14,19\}$ attain particularly low values.
Specifically, for $g=19$ this is no surprise since the prediction of the \ac{QNN} is based on the measurement of the first qubit such that the last rotation gate on the second qubit has no contribution to the prediction, \ie, the accuracy of the model.\par
\begin{figure}[!t]
	\centering
	\includefigure{\scaleA}{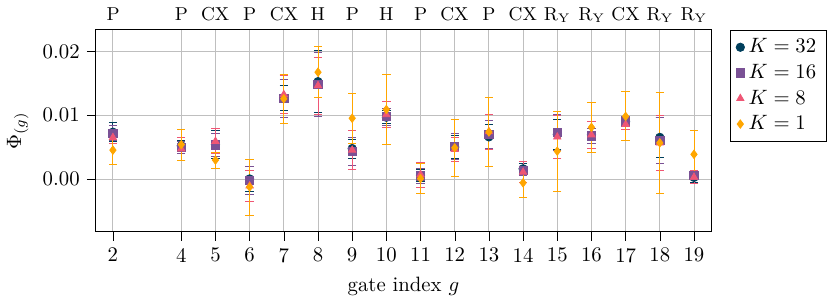}
	\caption{\Acp{QSV} $\Phi_{(g)}$, \cref{eqn:qphi}, for the feature map circuit of a \ac{QNN} in analogy to \cref{fig:qnn:results:phi} with the goal to compare different numbers of repetitions $K \in \{1,8,16,32\}$ to investigate the influence of shot noise.
We choose a sample fraction of $\alpha = \num{0.01}$, \cref{eqn:alpha}.
All evaluations are run on the \QCaer simulator.
Shown are mean and one standard deviation over five independent runs.}
	\label{fig:qnn:results:phiK}
\end{figure}

 
Our choice of value function and \ac{QPU} allows us to investigate the shot noise of this experiment.
For this purpose, we perform several additional runs with $K \in \{1,8,16,32\}$ repetitions.
The resulting \acp{QSV} are shown in \cref{fig:qnn:results:phiK}.\par
For $K=1$ the value function is based on a single one-shot prediction for each data point, which leads to a large uncertainty.
As expected, the shot noise becomes less significant for a higher number of repetitions $K$.
The plot indicates that between $K=16$ and $K=32$, no major improvement of the results can be observed, which implies that $K=16$ repetitions are already sufficient to capture the uncertainty of the shot noise in this experiment.

\subsection{Quantum generative adversarial network} \label{sec:qgan}
\begin{figure}[!t]
	\centering
	\includefigure{1}{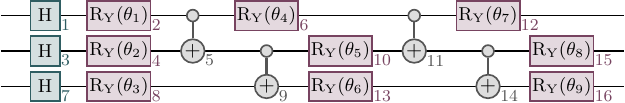}
	\caption{Circuit for the \ac{QGAN} with trainable parameters $\VEC{\theta} \in \mathbb{R}^9$.
		We use the circuit symbols from \cref{sec:app:circuit symbols} and the abbreviations from \cref{eqn:phixi,eqn:phix}.
		The number near each gate represents its gate index $g$, \cref{eqn:U}.}
	\label{fig:qgan:circuit}
\end{figure}

As third use case, we consider a \ac{QGAN} with the task of generating a discretized log-normal distribution as described in \cite{zoufal2019}.
For a recent review of \acp{QGAN} we refer to \cite{kiani2022}.
We use the circuit shown in \cref{fig:qgan:circuit} as our generative model.
It consists of three qubits and can therefore produce $2^3=8$ different amplitudes.
\par
\begin{figure}[!t]
	\centering
	\includefigure{1}{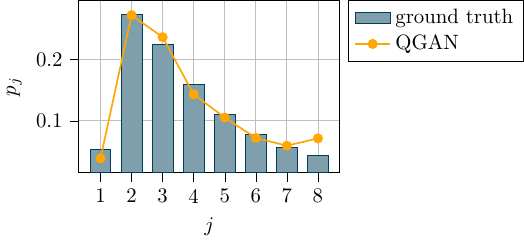}
	\caption{Output of the trained \ac{QGAN} and corresponding ground truth for the discretized log-normal distribution $p_j$ with a resolution of eight values $j$.
The \ac{QGAN} is evaluated on a \QCstatevector simulator.}
	\label{fig:qgan:data}
\end{figure}

First of all, we train the \ac{QGAN} by suitably optimizing its trainable parameters.
As a result, we find $\VEC{\theta} = \VEC{\theta}_{\mathrm{trained}} \approx ( \num{-1.328}, \num{-0.155}, \num{3.025}, \num{1.424}, \num{0.427}, \num{1.620}, \num{0.980}, \num{1.495}, \num{1.461} )$.
The \ac{QGAN} output in comparison with the ground truth is shown in \cref{fig:qgan:data}.
In the following, we only consider the trained \ac{QGAN}.
\par
As active gates for the evaluation of \acp{QSV}, we choose all gates except for the initial three $\HADAMARD$ gates $R=\{1,3,7\}$, \cref{eqn:R}, \ie, $A =\{ 2,4,5,6,8,9,\dots,16 \}$, \cref{eqn:A}.
For the value function $v(\subS)$, we choose the Hellinger fidelity, 
\cref{eqn:v:H}, that results from a comparison of the \ac{QGAN} output from the 
\QCstatevector simulator versus the \ac{QGAN} output when evaluated on the 
\ac{NISQ} \ac{QPU} \QCoslo.
The value function is consequently uncertain in the 
sense of \cref{eqn:v} and allows us to asses hardware errors.
For \QCoslo, we 
choose four different configurations.
Specifically, we map the logical qubits 
to the physical qubits of the backend in two different ways.
First, the three 
logical qubits of the QGAN circuit (from top to bottom in 
\cref{fig:qgan:circuit}) are mapped to the physical qubits 0-1-2 of \QCoslo.

\begin{figure}[!t]
	\centering
	\includefigure{\scaleB}{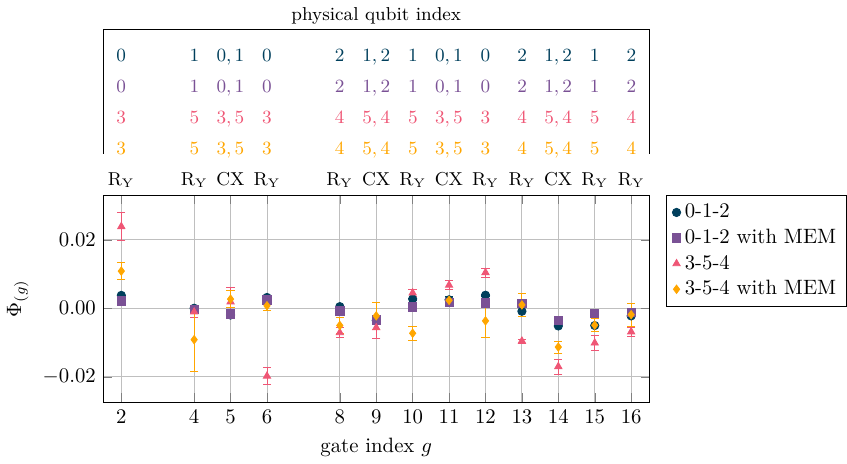}
	\caption{\Acp{QSV} $\Phi_{(g)}$, \cref{eqn:qphi}, for the \ac{QGAN} circuit from \cref{fig:qgan:circuit} with respect to a value function that corresponds to the Hellinger fidelity, \cref{eqn:v:H}, of the \ac{QGAN} output on the \QCstatevector simulator versus the \ac{QGAN} output on \QCoslo for different configurations.
First, using the physical qubits 0-1-2 and second, using the physical qubits 3-5-4, each with and without \ac{MEM}.
We show the mean and one standard deviation over three independent runs for each configuration.
A larger \ac{QSV} is more favorable.
For all instances, we choose $K=3$ and $\alpha=\num{1}$.
In analogy to the previous experiments, the gate index $g$, \cref{eqn:U}, is shown on the bottom, whereas the corresponding gate name is shown on the top of the plot.
In addition, we also list the physical qubit indices over each gate to indicate on which physical qubits it acts.
For the gates with $g \in R=\{1,3,7\}$, no \ac{QSV} is determined.}
	\label{fig:qgan:results:phi}
\end{figure}

\begin{figure}[!t]
	\centering
	\begin{subfigure}[t]{1.\linewidth}
		\centering
		\includefigure{1}{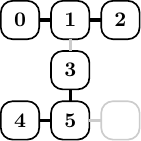}
		\caption{Connectivity map of \QCoslo, also see \cref{sec:app:backends}.}
		\label{fig:qgan:result:backend-oslo:connectivity}
	\end{subfigure}%
	\\[.25cm]
	\begin{subfigure}[t]{.49\linewidth}
		\centering
		\includefigure{\scaleA}{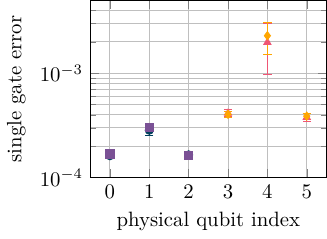}
		\caption{Single gate errors for \QCoslo as provided by IBM.}
		\label{fig:qgan:result:backend-oslo:e1}
	\end{subfigure}%
	\hfill	
	\begin{subfigure}[t]{.49\linewidth}
		\centering
		\includefigure{\scaleA}{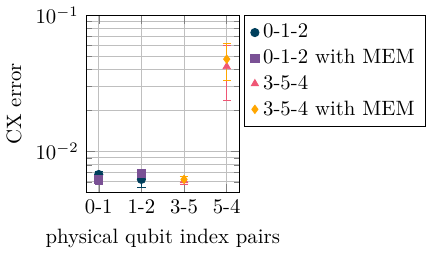}
		\caption{$\CNOT$ errors for \QCoslo as provided by IBM.}
		\label{fig:qgan:result:backend-oslo:e2}
	\end{subfigure}%	
	\caption{Gate errors as provided by IBM for \QCoslo (plotted with logarithmic scale).
In \subref{fig:qgan:result:backend-oslo:connectivity} we show the connectivity map of \QCoslo with physical qubit indices and connections between qubits that allow the realization of joint $\CNOT$ gates, where we highlight the two sets of physical qubits (0-1-2 and 3-5-4) that are used for the results in \cref{fig:qgan:results:phi}.
For each run from \cref{fig:qgan:results:phi}, we collect the respective errors that are provided by IBM \cite{ibmq2022a}.
Specifically, we fetch the error information for the hardware device every hour during the total runtime; also see \cref{sec:app:qgan timing}.
The results are presented in \subref{fig:qgan:result:backend-oslo:e1} and \subref{fig:qgan:result:backend-oslo:e2}, respectively, where we show the mean and one standard deviation over all error samples.} 
	\label{fig:qgan:result:backend-oslo}
\end{figure}

Second, the physical qubits 3-5-4 are chosen.
The two sets of physical qubits 
are also shown in \cref{fig:qgan:result:backend-oslo:connectivity}.
For each 
set, we perform an evaluation with and without \ac{MEM}.
The \ac{MEM} method we 
use here requires a preprocessing step (independent of the \ac{QGAN} circuit) 
in which the fraction of bit flips for each combination of measured ones and 
zeros is determined such that a linear transformation to any subsequent 
measurement result realizes the \ac{MEM} \cite{qiskit2021,nation2021}.
We 
perform three independent runs with $K=3$ and $\alpha=1$, \cref{eqn:alpha}.
The 
results are shown in \cref{fig:qgan:results:phi}.\par
Comparing the two instances with and without \ac{MEM}, we find that the chosen mitigation approach does not consistently increase the \acp{QSV} but may in some cases also lead to a decrease.
To further analyze our results, we require a frame of reference.
For this purpose, we can use the error information about the hardware device that is provided by IBM \cite{ibmq2022a}.
For technical reasons, these errors may change over time and are therefore updated regularly \cite{baheri2022}, which is why we collect a set of errors over the duration of the experiments.
Specifically, for each of the three runs from the four configurations shown in \cref{fig:qgan:results:phi}, we fetch the error information every hour during the total runtime (including waiting times and the execution of \ac{MEM}) as summarized in \cref{sec:app:qgan timing}.
The results are shown in \cref{fig:qgan:result:backend-oslo:e1,fig:qgan:result:backend-oslo:e2}.\par
Clearly, the physical qubits 0, 1, and 2 exhibit a smaller mean (and standard deviation) of the single gate errors than 3, 4, and 5.
Similarly, the $\CNOT$ errors of the physical qubit pairs 0-1, 1-2 and 3-5, show a considerably smaller mean (and standard deviation) than 5-4.
From these results one would expect that the configuration 0-1-2 should lead to significantly better results than the configuration 3-5-4.\par
If we compare these observations to our results from \cref{fig:qgan:results:phi}, it becomes apparent that the ordering of the errors provided by IBM is not always reflected by the ordering of the corresponding \acp{QSV}.
This can be seen for example when comparing $g=2$ with $g=6$ and $g=9$ with $g=14$, respectively.
This result might be a fact of the fundamental difference between the two error definitions: Our proposed error measure via \acp{QSV} is context-aware in the sense that it is defined for specific gates within a specific circuit, whereas the errors provided by IBM can in contrast be considered as generic.
A more detailed comparison is a possible starting point for further studies, which may also take into account recent error characterizations based on discrete time crystals that suggest that ``benchmarks computed infrequently are not indicative of the real-world performance of a quantum processor'' \cite{zhang2023}.

\subsection{Transpilation} \label{sec:classical preprocessing}
\begin{figure}[!t]
	\centering
	\begin{subfigure}[t]{1.\linewidth}
		\centering
		\includefigure{1}{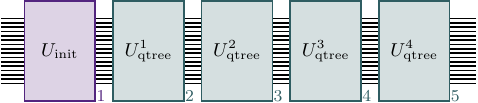}
		\caption{\Ac{QT} \cite{heese2022a} containing four subsequent branches, each represented by a decision layer $U^i_{\mathrm{qtree}}$.
The circuit operates on 16 qubits.
We choose $A =\{2,3,4,5\}$ and $R = \{1\}$.}
		\label{fig:tp:circuits:tree}
	\end{subfigure}%	
	\\[.25cm]
	\begin{subfigure}[t]{1.\linewidth}
		\centering
		\includefigure{1}{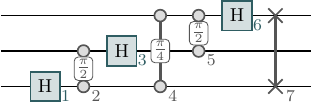}
		\caption{\Ac{QFT} for three qubits.
We choose $A=\{1,\dots,7\}$ and $R=\varnothing$.}
		\label{fig:tp:circuits:qft:3}
	\end{subfigure}%
	\\[.25cm]
	\begin{subfigure}[t]{1.\linewidth}
		\centering
		\includefigure{1}{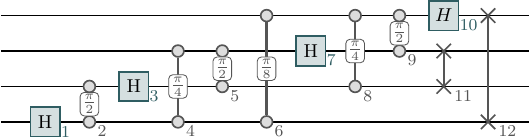}
		\caption{\Ac{QFT} for four qubits.
We choose $A=\{1,\dots,12\}$ and $R=\varnothing$.}
		\label{fig:tp:circuits:qft:4}
	\end{subfigure}%
	\\[.25cm]
	\begin{subfigure}[t]{1.\linewidth}
		\centering
		\includefigure{1}{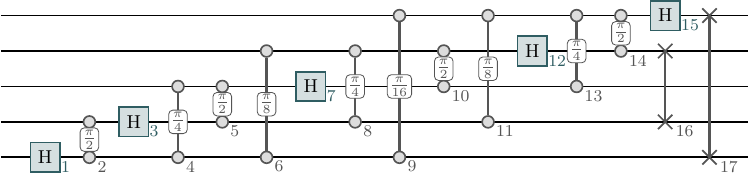}
		\caption{\Ac{QFT} for five qubits.
We choose $A=\{1,\dots,17\}$ and $R=\varnothing$.}
		\label{fig:tp:circuits:qft:5}
	\end{subfigure}%
	\caption{Circuits for the purpose of evaluating the estimated execution efficiency.
We specifically consider a \ac{QT} and three \acp{QFT} for a different number of qubits.
	We use the circuit symbols from \cref{sec:app:circuit symbols}.
The \ac{QT} circuit contains an initialization layer ($U_{\mathrm{init}}$) and decision layers ($U^i_{\mathrm{qtree}}$ for $i\in\{1,\dots,4\}$).
The number near each gate represents its gate index $g$, \cref{eqn:U}.
For each circuit, we list the choice of active gates $A$, \cref{eqn:A}, and remaining gates $R$, \cref{eqn:R}.}
	\label{fig:tp:circuits}
\end{figure}

The fourth use case addresses circuit transpilation as a form of classical preprocessing.
Specifically, we consider the transpilation of two types of circuits:
\begin{itemize}
	\item A \ac{QT} with four decision branches trained on the binarized tic-tac-toe endgame data set \cite{aha2021} as described in \cite{heese2022a}.
The model requires 16 qubits.
For the implementation, we use \cite{heese2022git}.
	\item \Acp{QFT} \cite{coppersmith2002} for three to five qubits.
\end{itemize}
The respective circuits are shown in \cref{fig:tp:circuits}, where we also list the choice of active gates $A$, \cref{eqn:A}, and remaining gates $R$, \cref{eqn:R}, respectively.
A \ac{QT} consists of an initialization layer and subsequent decision layers that represent the binary decisions within the tree model from the root to the branches.
\par
\begin{figure}[!t]
	\centering
	\begin{subfigure}[t]{.29\linewidth}
	\centering
	\includefigure{\scaleB}{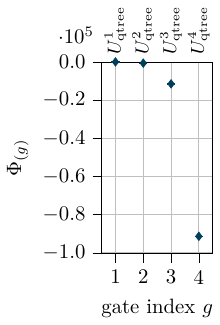}
	\caption{\Ac{QT}}
	\label{fig:tp:results:phi:qtree}
	\end{subfigure}%	
	\hfill
	\begin{subfigure}[t]{.69\linewidth}
	\centering
	\includefigure{\scaleB}{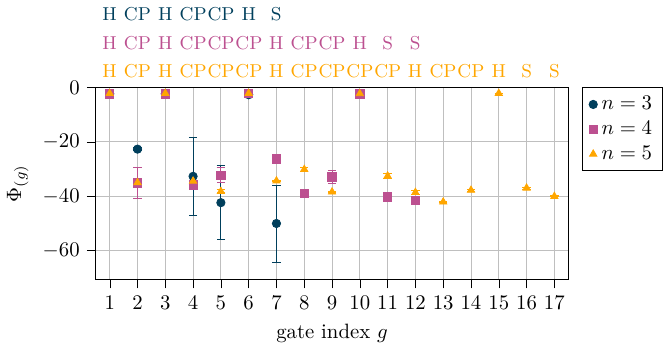}
	\caption{\Ac{QFT} for $n$ qubits}
	\label{fig:tp:results:phi:qft}
	\end{subfigure}%
	\caption{\Acp{QSV} $\Phi_{(g)}$, \cref{eqn:qphi}, for the \ac{QT} and \ac{QFT} circuits, respectively, from \cref{fig:tp:circuits} with a value function that corresponds to the transpilation-based estimated execution efficiency, \cref{eqn:v:transpilation}, with respect to the \QCehningen \ac{QPU}.
A larger \ac{QSV} is more favorable.
In analogy to the previous experiments, the gate index $g$, \cref{eqn:U}, is shown on the bottom, whereas the corresponding gate (or layer) name is shown on the top of the plot.
All evaluations are run on a classical host with $K=1$.
We choose $\alpha=\num{1}$ for \ac{QT} and $\alpha=\num{0.01}$ for \ac{QFT}, respectively.
In total, we perform three independent runs and show the mean value and one standard deviation (not visible in \subref{fig:tp:results:phi:qtree}).}
	\label{fig:tp:results:phi}
\end{figure}

As value function $v(\subS)$, we choose the transpilation-based estimated execution efficiency, \cref{eqn:v:transpilation}, with penalty parameters $s_1=-1$ and $s_2=-10$.
Its evaluation is performed exclusively on a classical host, where we repeat the transpilation process \num{50} times and take the best result.
No measurements of the circuits take place for the evaluation.
Transpilation is performed with respect to the \QCehningen \ac{QPU} using the Qiskit transpiler.
Randomness is introduced by the heuristic transpilation algorithm that leads to an uncertain value function in the sense of \cref{eqn:v}.
This is mitigated, to some extend, by our proposed approach of selecting the best transpilation result out of multiple repetitions.
We choose $K=1$ as well as $\alpha=\num{1}$ for \ac{QT} and $\alpha=\num{0.01}$ for \ac{QFT}, \cref{eqn:alpha}.
In total, we perform three independent runs for each \ac{QFT} and the \ac{QT} circuit, respectively, with random sampling of value functions and random transpiler seeds.
The resulting \acp{QSV} are shown in \cref{fig:tp:results:phi}.
\par 
The \ac{QT} results indicate, as expected, that deeper decision tree layers have smaller \acp{QSV} since they involve an exponentially growing number of branches and therefore become increasingly more complex.
Specifically, the plot suggests an approximately exponential decrease of \acp{QSV} for deeper layers.
For the \ac{QFT}, we find that specific $\HADAMARD$ and $\CPHASE$ gates for $g\in\{1,3,6,10,15\}$ exhibit larger \acp{QSV}.
On the other hand, all remaining gates (which particularly includes all $\SWAP$ gates) show smaller \acp{QSV}.
Interestingly, for $n=3$ the \acp{QSV} of the gates with $g \in \{4,5,7\}$ have a comparably large standard deviation, whereas the standard deviation of almost all other gates is negligible.

\subsection{Variational quantum eigensolver} \label{sec:qaoa}
\begin{figure}[!t]
	\centering
	\includefigure{1}{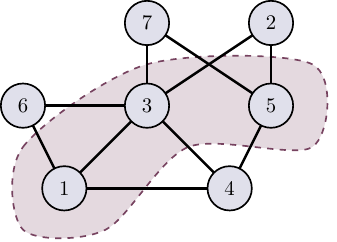}
	\caption{Exemplary graph for the max-cut problem consisting of seven vertices (each labeled by its unique index) and ten edges.
		The dashed line represents the optimal cut that leads to the unique solution to the problem in form of two subgraphs that are connected by nine edges.}
	\label{fig:maxcut-qaoa:graph}
\end{figure}

\begin{figure}[!t]
	\centering
	\includefigure{1}{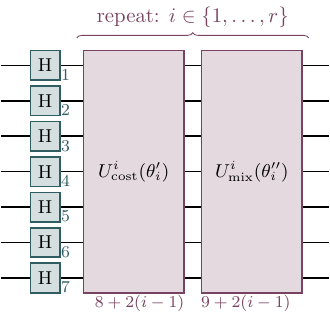}
	\caption{Circuit for \ac{QAOA} of depth $r$, which is parameterized by the parameters $\VEC{\theta} \in \mathbb{R}^{2r}$.
This circuit contains $\HADAMARD$ gates, a cost layer $U^i_{\mathrm{cost}}(\theta'_{i})$ depending on the parameter $\theta'_{i} := \theta_{2(i-1)+1} \in \mathbb{R}$, and a mixing layer $U^i_{\mathrm{mix}}(\theta''_{i})$ depending on the parameter $\theta''_{i} :=\theta_{2(i-1)+2} \in \mathbb{R}$.
The number near each gate represents its gate index $g$, \cref{eqn:U}.}
	\label{fig:maxcut-qaoa:circuit}
\end{figure}

In this final use case, we empirically investigate the usage of \acp{QSV} for a \ac{VQE} to solve an optimization task that is not from the domain of \ac{QML}.
Specifically, we consider a max-cut problem for the exemplary graph shown in \cref{fig:maxcut-qaoa:graph}.
This binary optimization problem has the goal of finding a bipartite subgraph of the original graph with as many edges as possible.
It is NP-hard by definition and therefore no polynomial-time algorithms are known.
We use the \ac{QAOA} to solve this problem on a \ac{QPU} \cite{farhi2014,moll2018,hadfield2019}.
The respective circuit is shown in \cref{fig:maxcut-qaoa:circuit} and depends on the chosen depth $r$ that defines the number of repetitions for cost layers (or cost unitaries) and mixing layers (or mixing unitaries).
By design, each layer consists of multiple gates, but we do not consider the internal structure for this experiment.
\par
As active gates for the evaluation of \acp{QSV}, we choose all problem and mixing layers.
Thus, the active gate set is formally given by $A =\{ 8+2(i-1)+k \,|\, i \in r \,\land\, k \in \{0,1\} \}$, \cref{eqn:A}, whereas the set of remaining gates reads $R=\{1,\dots,7\}$, \cref{eqn:R}.
We consider \ac{QAOA} circuits of different depth $r \in \{1,\dots,7\}$.
To evaluate \acp{QSV}, we first solve the optimization problem with a COBYLA 
optimizer for each depth to obtain parameters $\VEC{\theta} = 
\VEC{\theta}_{\mathrm{optimal}}$.
To investigate the generality of the results, 
we perform three independent optimization runs in total for each depth with 
different random seeds.
That is, we obtain three different sets of optimal 
parameters for each depth.
The resulting optimization goals are listed in 
\cref{sec:app:qaoa optimization results}.
Based on these fixed parameters, we 
evaluate the \acp{QSV} for the respective circuits.
In total, we consider three 
different value functions $v(\subS)$ for this purpose: (i) expressibility, 
\cref{eqn:v:expressibilty}, (ii) entangling capability, 
\cref{eqn:v:entangling}, and (iii) energy corresponding to the expectation 
value of the cost Hamiltonian
\begin{equation} \label{eqn:v:energy}
	v(\subS) = \braket{\Psi(\subS, R, \VEC{\theta}) | \mathcal{H} | \Psi(\subS, R, 
	\VEC{\theta})},
\end{equation}
where we recall \cref{eqn:psiUSR}.
The problem Hamiltonian $\mathcal{H}$ is defined in the usual way such that its ground state encodes the solution to the proposed max-cut problem.\par
\begin{figure}[p]
	\centering
	\begin{subfigure}[t]{1.\linewidth}
		\centering
		\includefigure{1}{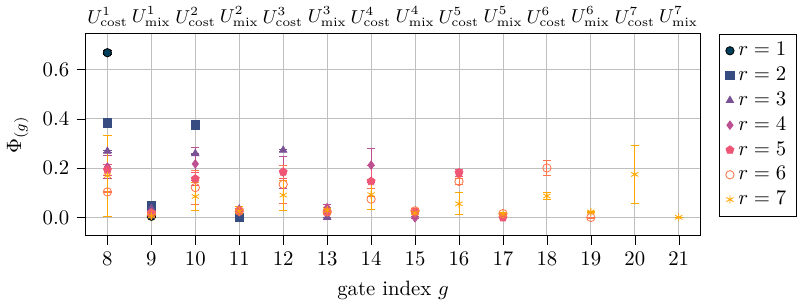}
		\caption{entangling capability}
		\label{fig:maxcut-qaoa:results:phi:ent}
	\end{subfigure}%
	\\[.25cm]
	\begin{subfigure}[t]{1.\linewidth}
		\centering
		\includefigure{1}{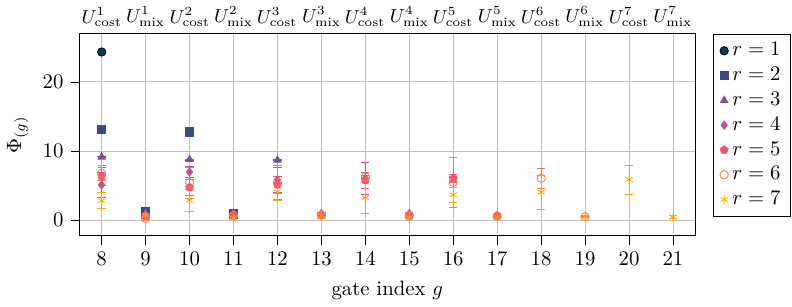}
		\caption{expressibility}
		\label{fig:maxcut-qaoa:results:phi:exp}
	\end{subfigure}%
	\\[.25cm]
	\begin{subfigure}[t]{1.\linewidth}
		\centering
		\includefigure{1}{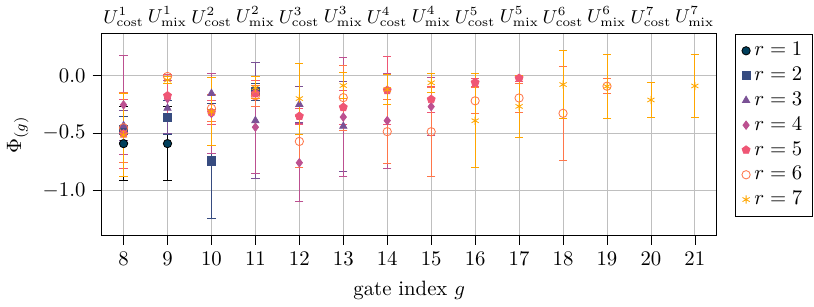}
		\caption{energy}
		\label{fig:maxcut-qaoa:results:phi:h}
	\end{subfigure}%
	\caption{\Acp{QSV} $\Phi_{(g)}$, \cref{eqn:qphi}, for the \ac{QAOA} circuit of depth $r \in \{1,\dots,7\}$, \cref{fig:maxcut-qaoa:circuit}, with three value functions of interest: \subref{fig:maxcut-qaoa:results:phi:ent} entangling capability, \cref{eqn:v:entangling}, \subref{fig:maxcut-qaoa:results:phi:exp} expressibility, \cref{eqn:v:expressibilty}, and \subref{fig:maxcut-qaoa:results:phi:h} energy, \cref{eqn:v:energy}.
For \subref{fig:maxcut-qaoa:results:phi:ent} and \subref{fig:maxcut-qaoa:results:phi:exp}, larger \acp{QSV} are more favorable, whereas smaller \acp{QSV} are more favorable for \subref{fig:maxcut-qaoa:results:phi:h}.
In analogy to the previous experiments, the gate index $g$, \cref{eqn:U}, is shown on the bottom, whereas the corresponding gate (\ie, layer) name is shown on the top of the plot.
For the gates with $g \in R=\{1,\dots,7\}$, no \ac{QSV} is determined.
All evaluations are run on the \QCstatevector simulator.
We choose $K=1$ and $\alpha=\num{1}$ for $r \in \{1,2,3\}$, $\alpha=\num{0.1}$ for $r \in \{4,5\}$, $\alpha=\num{0.01}$ for $r = 6$, and $\alpha=\num{0.001}$ for $r = 7$.
For each depth $r$, we show the mean and one standard deviation over three independent runs.}
	\label{fig:maxcut-qaoa:results:phi}
\end{figure}

We run all evaluations on the \QCstatevector simulator and therefore no randomness is involved in the evaluation of the value functions.
We choose $K=1$ and $\alpha=\num{1}$ for $r \in \{1,2,3\}$, $\alpha=\num{0.1}$ for $r \in \{4,5\}$, $\alpha=\num{0.01}$ for $r = 6$, and $\alpha=\num{0.001}$ for $r = 7$, \cref{eqn:alpha}.
For each repetition number, we conduct three independent runs with different random seeds for the algorithm optimizer as well as a different sample in the sense of \cref{eqn:phi:multisample}.
The resulting \acp{QSV} are presented in \cref{fig:maxcut-qaoa:results:phi}.
\par
The \acp{QSV} for the expressibility and entangling capability both reveal a similar structure, which clearly shows that cost layers have a dominating contribution, whereas mixing layers have a vanishing contribution.
For the energy, no clear structure of the respective \acp{QSV} can be identified and, in addition, the standard deviations over different independent optimization runs are also relatively large.
Hence, we can conclude that all gates attribute similarly to the expectation value independent of the depth and their specific contribution might vary depending on the optimized parameters.
This in particular means that the mixing layers contribute similarly to cost layers and therefore, pruning mixing layers from the circuit is not expected to bring a general benefit of the algorithm as can be expected from the theory.
Summarized, we find from our observations that the exploration of the unitary space with entangled qubits---\ie, the search space---is entirely driven by the cost layers (as the informative part of the algorithm containing the problem structure), whereas the mixing layers (as the uninformative part containing a generic structure) help to minimize the expectation value without contributing to an extension of the search space.
Based on these results one could also argue that expressibility and entangling capability are both not necessarily required for energy minimization.
And indeed, high expressibility can lead to flatter cost landscapes that hamper optimization \cite{holmes2022}.\par
\begin{figure}[!t]
	\centering
	\begin{subfigure}[t]{.49\linewidth}
		\centering
		\includefigure{\scaleB}{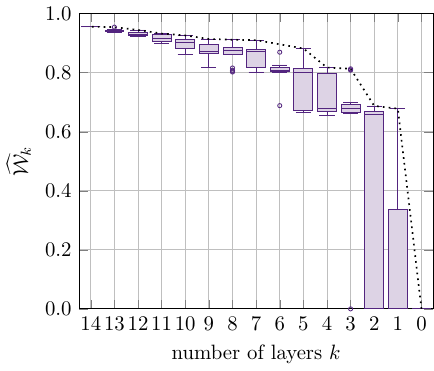}
		\caption{entangling capability}
		\label{fig:maxcut-qaoa:results:mem:ent}
	\end{subfigure}%	
	\hfill
	\begin{subfigure}[t]{.49\linewidth}
		\centering
		\includefigure{\scaleB}{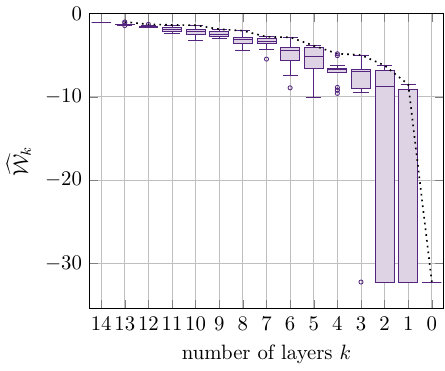}
		\caption{expressiblity}
		\label{fig:maxcut-qaoa:results:mem:exp}
	\end{subfigure}%
	\caption{Distribution of sampled value functions $\widehat{\mathcal{W}}_k$, \cref{eqn:W}, for different number of layers $k$ in form of box plots in analogy to \cref{fig:kernel-svm:results:mem}.
The results are obtained from one optimization run from \cref{fig:maxcut-qaoa:results:phi} with depth $r=7$.
All evaluations are run on a \QCstatevector simulator with $K=1$ and $\alpha=\num{0.001}$.
The dotted line indicates an approximation of the Pareto frontier for a multi-criteria circuit pruning problem with the goal to minimize the number of layers and to maximize the corresponding value function (expressibility and entangling capability, respectively).} 
	\label{fig:maxcut-qaoa:results:mem}
\end{figure}

To further investigate the importance of the circuit depth for the entangling capability and expressibility, we conduct a brief analysis of the sampled value function distribution depending on the number of active gates (\ie, layers) according to \cref{eqn:W} in analogy to \cref{fig:kernel-svm:results:mem}.
For this purpose, we limit ourselves to the case $r=7$.
In this analysis, we do not consider the energy function since (in contrast to the entangling capability and the expressibility) it depends on the optimization parameters which have been determined beforehand based on the complete circuit such that starting with a pruned circuit might converge to a different set of optimization parameters with a different energy.
The results are shown in \cref{fig:maxcut-qaoa:results:mem}.\par
As expected, we find that there is a clear tradeoff between the number of layers and the expressibility and entangling capability, respectively.
For a smaller number of layers, the distribution also becomes significantly broader.

%% file: conclusions.tex
\section{Conclusions} \label{sec:conclusions}
In this work, the important \ac{XAI} tool of \acfp{SV} has been adopted to explain quantum circuits by analyzing the influence of gates (or groups of gates) on quantum computations. We use the term \acfp{QSV} for our approach. While we have mostly focused on its use for approaching model architecture investigations in an \ac{XQML} context, our method is in fact applicable beyond \ac{QML} to gate-based quantum computations in general.\par
The value function for the evaluation of \acp{QSV} can be freely chosen, which allows us to study the impact of quantum gates on a wide range of properties, also within quantum-classical hybrid systems such as a quantum-enhanced \ac{ML} pipeline. We demonstrate this versatility experimentally with five detailed use cases including \ac{QML} models for classification and generative modeling as well as quantum optimization tasks with variational quantum circuits. Within these use cases, we perform various explanations with different value functions ranging from model performance over hardware noise and transpilation efficiency to circuit expressibility and entanglement capability. To realize the experiments, we have used both simulators running on a classical host and two different \acp{QPU} based on superconducting quantum hardware.\par
Given this broad applicability, our method opens up opportunities for monitoring and understanding the behavior of real-world quantum computing systems.
At the same time, new research questions emerge, \eg, understanding the relation between qubit error metrics or crosstalk \cite{sarovar2020} and the context-aware (\ie, gate-specific and circuit-specific) quantities derived via \acp{QSV} to quantify hardware uncertainties for generative models.
In addition, our findings also indicate that \ac{QSV} could be a promising tool within a \ac{QCAS} framework.\par
The high complexity, inherited from the classical \acp{SV}, is arguably the major limitation of the proposed \ac{QSV} approach.
A high computational effort is required to cover relevant coalitions and to reduce the uncertainty of the estimated quantities.
While our results suggest that an estimation with a subset of coalitions on actual \acp{QPU} delivers reasonable results, further research might be required to allow a more general analysis, especially for large scale systems.\par
Moreover, \acp{QSV} could be modified or extended in many ways as a possible starting point for further studies:
\begin{itemize}
	\item In place of \acp{SV}, Owen values \cite{owen1977,lopez2009} or Banzhaf values \cite{banzhaf1965} can be used, concepts that are already used in classical \ac{XAI} \cite{chen2020,karczmarz2021}.
	\item Instead of removing gates that are not part of the coalition, other operations could be performed on those gates, \eg, local inversions as proposed recently to quantify the contribution of individual gates to the overall error on \ac{NISQ} hardware \cite{calderonvargas2023}.
	\item Certain aspects of \ac{XQML} like, \eg, an intuitive explanation of quantum feature spaces \cite{lloyd2020}, go beyond the scope of classical \ac{XAI} and might require entirely new explainability approaches.
\end{itemize}
These exemplary aspects indicate that there are still many open research aspects for the explanation of quantum circuits, even within the limited scope of our approach.
\par
So far, \ac{XQML} is a novel, yet mostly unexplored field that can in the spirit of \cite{schuld2022} be considered as an alternative research direction of \ac{QML} instead of the search for a quantum advantage. In this sense, we see this contribution as one of the first shots at \ac{XQML} with a focus on \ac{QML} model architecture and as a promising cornerstone for further research.

%% file: acknowledgments.tex
Parts of this research have been funded by the Ministry of Science and Health of the State of Rhineland-Palatinate as part of the project \emph{AnQuC}.
Parts of this research have been funded by the Federal Ministry of Education and Research of Germany and the state of North-Rhine Westphalia as part of the \emph{Lamarr-Institute for Machine Learning and Artificial Intelligence}.
Access to quantum computing hardware has been funded by \emph{Fraunhofer Quantum Now}.

%% file: appendix.tex
\appendixsection{Expressibility} \label{sec:app:expressibility}
In this appendix section, we briefly present an estimator for the expressibility along the lines of \cite{sim2019} with the minor extension that we focus on a formulation that can actually be measured on a \ac{QPU}. For this purpose, we consider an ansatz circuit consisting of $q$ qubits and a corresponding unitary, \cref{eqn:psiUSR}, that is fully determined by a parameter vector $\VEC{\theta} \in \mathcal{P}$, whereas a constant feature vector $\VEC{x} \in \mathcal{X}$ (\ie, $\VEC{x} = \const$) is presumed. An estimator of the expressibility of this circuit is given by the Kullback-Leibler divergence
\begin{equation} \label{eqn:app:expr}
	\eta(\subS,R,\VEC{x},c_p,c_b,\mathcal{F}) := 
	D_{\mathrm{KL}}[\hat{P}^{c_p,c_b}_{\mathrm{circ}}(F,\subS,R,\VEC{x},\mathcal{F})
	 \Vert P_{\mathrm{Haar}}(F)]_F \geq 0
\end{equation}
with respect to the fidelity $F \in [0,1]$.\par
The first probability distribution in \cref{eqn:app:expr} is an estimation of the probability distribution of fidelities for the circuit of interest. It can be obtained by introducing two \iid random variables $\VEC{\RV{\vartheta}}_1 \sim u(\mathcal{P})$ and $\VEC{\RV{\vartheta}}_2 \sim u(\mathcal{P})$, where $u(\mathcal{P})$ denotes the uniform distribution on $\mathcal{P}$. These random variables represent randomly chosen parameter vectors. In total, $c_p$ pairs $\{(\VEC{\vartheta}_1^1, \VEC{\vartheta}_2^1),\dots,(\VEC{\vartheta}_1^{c_p}, \VEC{\vartheta}_2^{c_p})\}$ of those random variables are sampled. For each realization pair $i \in \{1,\dots,c_p\}$, a SWAP test \cite{barenco1997,buhrmann2001} is performed based on $s$ shots.\par
From these measurements, we obtain the set of bit strings
\begin{equation}
	\VEC{B}^i(\VEC{x}, \VEC{\vartheta}_1^i, \VEC{\vartheta}_2^i) \equiv \VEC{B}^i := \{ b^i_1, \dots, b^i_s \}
\end{equation}
in analogy to \cref{eqn:B} with $b^i_j \in \{0,1\} \,\forall\, j \in \{1,\dots,s\}$. Furthermore, $b^i_j \sim p_{\mathrm{SWAP}}(b^i_j)$ with
\begin{equation}
	p_{\mathrm{SWAP}}(b^i_j=1) = \frac{1}{2} - \frac{1}{2} F^i,
\end{equation}
where
\begin{equation}
	F^i \equiv F^i(\subS, R, \VEC{x}, \VEC{\vartheta}_1^i, \VEC{\vartheta}_2^i) 
	:= \vert\braket{\Psi(\subS, R, \VEC{x}, \VEC{\vartheta}_1^i) | \Psi(\subS, 
	R, \VEC{x}, \VEC{\vartheta}_2^i)}\vert \in [0,1]
\end{equation}
denotes the fidelity of the two states resulting from the circuit of interest for the randomly sampled parameter vector pair $i$, where we recall \cref{eqn:psiU}. Thus, an estimate
\begin{equation}
	\hat{F}^i := 1 - 2 \frac{\vert \{ b \,|\, b \in \VEC{B}^i, b = 1 \} \vert }{s}
\end{equation}
of $F^i$ can be obtained from the multiset of bit strings.\par
In total, $s c_p$ shots are necessary to measure the set of fidelities $\mathcal{F} := \{\hat{F}^1,\dots,\hat{F}^{c_p}\}$ for all pairs. From a histogram of $\mathcal{F}$ based on $c_b$ evenly sized bins, the probability distribution
\begin{align}
	\hat{P}^{c_p,c_b}_{\mathrm{circ}}(F,\subS,R,\VEC{x},\mathcal{F}) := 
	\frac{1}{c_p} \Bigg\vert \Bigg\{ F' \,|\, & F' \in \mathcal{F}, \left[ 
	\frac{j(c_b,F)}{c_b} \leq F' < \frac{j(c_b,F)+1}{c_b} \right] \nonumber\\
	& \lor \left[ j(c_b,F)+1 = c_b \land F' = 1\right] \Bigg\} \Bigg\vert^2
\end{align}
with
\begin{equation}
	j(c_b,F) := \begin{cases} \lfloor F c_b \rfloor & \text{if}\,\, 0 \leq F < 1 \\ c_b-1 & \text{if}\,\, F = 1 \end{cases}
\end{equation}
can be obtained.\par
The second probability distribution in \cref{eqn:app:expr} represents the probability density function of fidelities for the ensemble of Haar random states with the analytical form
\begin{equation}
	P_{\mathrm{Haar}}(F) := (2^q-1)(1-F)^{2^q-2},
\end{equation}
which only depends on the number of qubits $q$.\par
The presented expressibility measure, \cref{eqn:app:expr}, is based on a distance measure such that a high value indicates a low expressibility and vice versa. For a more detailed discussion, we refer to \cite{sim2019,harrow2009,nakaji2021} and references therein. An alternative measure can also be found in \cite{du2022a}.

\appendixsection{Entangling capability} \label{sec:app:entangling capability}
This appendix section outlines a measure for the entangling capability following \cite{sim2019}. We extend the original concept with a brief description of how it can be measured on a \ac{QPU} by means of quantum tomography. In analogy to the expressibility calculation in \cref{sec:app:expressibility}, we consider an ansatz circuit consisting of $q$ qubits and a corresponding unitary, \cref{eqn:psiUSR}, that is fully determined by a parameter vector $\VEC{\theta} \in \mathcal{P}$, whereas a constant feature vector $\VEC{x} \in \mathcal{X}$ (\ie, $\VEC{x} = \const$) is presumed. A measure for the entangling capability of this circuit is given by the mean
\begin{equation} \label{eqn:app:ent}
	\lambda(\subS,R,\VEC{x},c_s,\mathcal{T}) := \frac{1}{c_s} \sum_{i=1}^{c_s} 
	Q( \ket{\hat{\Psi}^i} ) \in [0,1]
\end{equation}
over the Meyer-Wallach entanglement measure \cite{meyer2002}, that is defined as 
\begin{equation}
	Q(\ket{\Psi}) := \frac{4}{n} \sum_{j=1}^q D( \iota_j(0) \ket{\Psi}, \iota_j(1) \ket{\Psi} ) \in [0,1]
\end{equation}
for a $q$-qubit state $\ket{\Psi}$. Here, we use the mapping
\begin{equation}
	\iota_j(b) \ket{b_1 \cdots b_q} := \delta_{b b_j} \ket{b_1 \cdots \hat{b}_j \cdots b_q}
\end{equation}
that acts on the computational basis, where $\hat{b}_j$ denotes the absence of the $j$th qubit and $b \in \{0,1\}$. Furthermore, we make use of the generalized distance
\begin{equation}
	D( \ket{u}, \ket{v} ) := \frac{1}{2} \sum_{i,j=1}^{2^q-2} \vert u_i v_j  - u_j v_i \vert^2
\end{equation}
with $\ket{u} := \sum_{i=1}^{2^q-2} u_i \ket{i}$ and $\ket{v} := \sum_{i=1}^{2^q-2} v_i \ket{i}$, respectively.\par
In \cref{eqn:app:ent}, the Meyer-Wallach entanglement measure is evaluated for all quantum states in $\mathcal{Q} := \{\ket{\hat{\Psi}^i},\dots,\ket{\hat{\Psi}^{c_s}}\}$. To determine these states, a random variable $\RVEC{\vartheta} \sim u(\mathcal{P})$ is introduced and $c_s$ realizations $\mathcal{T}:=\{\VEC{\vartheta}^1\dots,\VEC{\vartheta}^{c_s}\}$ are obtained. For each realization $i \in \{1,\dots,c_s\}$, a quantum state tomography \cite{zambrano2020} is performed to obtain
\begin{equation}
	\ket{\hat{\Psi}^i} \equiv \ket{\hat{\Psi}^i(\subS, R, \VEC{x}, 
	\VEC{\vartheta}^i, \VEC{B}^i)} \approx \ket{\Psi(\subS, R, \VEC{x}, 
	\VEC{\vartheta}^i)},
\end{equation}
where we recall \cref{eqn:psiUSR}. We presume that a multiset
\begin{equation}
	 \VEC{B}^i \equiv \VEC{B}^i(\subS, R, \VEC{x}, \mathcal{T}) := \{ b^i_1, 
	 \dots, b^i_s \}
\end{equation}
of $s$ measurement results in form of bit strings in analogy to \cref{eqn:B} with $b^i_j \in \{0,1\} \,\forall\, j \in \{1,\dots,s\}$ is necessary to achieve this goal, but do not further specify the tomographic process
\begin{equation}
	\VEC{B}^i \mapsto \ket{\hat{\Psi}^i},
\end{equation}
which can for example correspond to the iterative algorithm presented in \cite{zambrano2020}.
In total, $s c_s$ shots are necessary to estimate all quantum states of interest.\par
According to the Meyer-Wallach entanglement measure, the entangling capability, \cref{eqn:app:ent}, becomes larger for more entangled states in $\mathcal{T}$. For a more detailed discussion, we refer to \cite{sim2019,meyer2002} and references therein. An alternative interpretation and evaluation of \cref{eqn:app:ent} is presented in \cite{brennen2003}.

\appendixsection{Hellinger fidelity} \label{sec:app:hellinger fidelity}
In this appendix section, we briefly outline a metric to quantify \ac{NISQ} 
hardware deficiencies. For this purpose, we consider a circuit of interest 
consisting of $q$ qubits and a corresponding unitary, \cref{eqn:psiUSR}, that 
may optionally depend on a parameter vector $\VEC{\theta} \in \mathcal{P}$ and 
feature vector $\VEC{x} \in \mathcal{X}$, respectively, which are both presumed to be constant (\ie, $\VEC{\theta} = \const$ and $\VEC{x} = \const$). The circuit is 
evaluated independently with two different approaches. First, on a \ac{QPU} to 
obtain the measurement results $\VEC{B}(\subS)$, \cref{eqn:BSR}, as a 
collection of bit strings. And second, with an idealized (\ie, noise-free) 
simulator running on classical hardware.\par
The estimated probability distribution of bit strings $\VEC{b} \in \{0,1\}^{q}$ that can be obtained from the \ac{QPU} is given by
\begin{equation}
	m_{\mathrm{qc}}(\VEC{b},\VEC{B}(\subS)) := \frac{\vert \{ \VEC{b}' \,|\, 
	\VEC{b}' \in \VEC{B}(\subS), \VEC{b}' = \VEC{b} \} \vert}{s}
\end{equation}
for $s$ shots. On the other hand, the corresponding probability distribution obtained from the classical simulator is given by $m_{\mathrm{sim}}(\VEC{b}) \equiv m_{\mathrm{sim}}(\VEC{b}, S, R, \VEC{x}, \VEC{\theta})$ according to \cref{eqn:bSR}. The Hellinger fidelity \cite{hellinger1909}
\begin{equation} \label{eqn:app:fidelity}
	H(\subS, R, \VEC{x}, \VEC{\theta}, \VEC{B}(\subS)) := \left[\sum_{\VEC{b} 
	\in \{0,1\}^{q}} \sqrt{m_{\mathrm{qc}}(\VEC{b},\VEC{B}(\subS)) 
	m_{\mathrm{sim}}(\VEC{b})} \right]^2 \in [0,1]
\end{equation}
is a similarity measure of the two distributions. It can therefore be used as a metric to quantify \ac{NISQ} hardware deficiencies. By definition, a smaller fidelity indicates larger hardware deficiencies.

\appendixsection{Estimated execution efficiency} \label{sec:app:estimated execution efficiency}
This appendix section considers the estimated execution efficiency as a measure of how efficiently a circuit can be executed on a specific \ac{NISQ} hardware device. The challenge of efficient execution arises because of two typical limitations of \ac{NISQ} hardware. First, only a limited set of gates can be physically realized on a hardware device. However, presuming that this set of available gates is a universal set of quantum gates, all other gates can be expressed in terms of the available gates \cite{nielsen2011,tucci2018}. Second, the connectivity of physical qubits in a hardware device is limited in the sense that gates acting on two (or more) qubits can only be applied on certain pairs (or sets) of qubits. However, this limitation can be overcome by introducing additional gates to swap qubits accordingly \cite{nielsen2011}. Due to these two limitations, any quantum circuit must be subjected to a suitable equivalence transformation before execution such that it contains only gates from the set of available gates $\mathcal{U}$ as defined by the hardware device of interest. This equivalence transformation is also called transpilation \cite{qiskit2021}.\par 
Specifically, we consider a transpilation procedure $T$ as an equivalence 
transformation of a gate set $P(\subS,R) := \{ A_a \,|\, a \in S\} \cup R$ from 
\cref{eqn:USR} into a new gate set $P'(\subS,R) := \{1,\dots,G'\} \cup R$ 
according to
\begin{equation}
	T: P(\subS,R) \mapsto P'(\subS,R)
\end{equation}
such that
\begin{equation}
	U(\subS, R, \VEC{x}, \VEC{\theta}) = U'(\subS, R, \VEC{x}, \VEC{\theta}),
\end{equation}
where
\begin{equation}
	U'(\subS, R, \VEC{x}, \VEC{\theta}) := \prod_{g \in P'(\subS,R)} 
	U_g(\VEC{x}^g, \VEC{\theta}^g)
\end{equation}
in analogy to \cref{eqn:USR} with unitary operators $U'_g(\VEC{x}^g, 
\VEC{\theta}^g) \in \mathcal{U}$ for all $g \in P'(\subS,R)$. The set 
$\mathcal{U}$ contains all gates that are available on the hardware device of 
interest. For the sake of simplicity, we presume that $\mathcal{U}$ contains 
only one-qubit and two-qubit gates.\par
Generally, the equivalence transformation of a circuit is ambiguous and can therefore lead to different gate decompositions with a varying number of gates. Since every gate introduces hardware-related uncertainty into the corresponding \ac{QPU} computation, an efficient transpilation with at least gates as possible in the resulting circuit is desired, which is a non-trivial optimization task. The estimated transpilation efficiency
\begin{equation} \label{eqn:app:tauT}
	\tau_T(\subS, R, \mathcal{U}, s_1, s_2) := \sum_{g \in P'(\subS,R)} 
	\hat{\tau}( U'_g(\VEC{x}^g, \VEC{\theta}^g), s_1, s_2) \leq 0
\end{equation}
is a possible metric to measure the quality of a certain transpilation $T$ for an available gate set $\mathcal{U}$ based on the resulting gate decomposition. The penalty function
\begin{equation}
	\hat{\tau}(U, s_1, s_2) := \begin{cases} s_2 & \text{if}\,\, U \in \mathcal{U}_{\mathrm{2}} \\ s_1 & \text{otherwise} \end{cases}
\end{equation}
estimates the decrease in efficiency for a higher number of gates by assigning a negative penalty value to each gate, where $\mathcal{U}_{\mathrm{2}} \subset \mathcal{U}$ contains all available two-qubit gates. Hence, a higher penalty score indicates a higher estimated transpilation efficiency (since less noise can be expected). The penalty parameters $s_1, s_2 \in \mathbb{R}_{<0}$ can be chosen arbitrarily, where typically $0 > s_1 > s_2$ to reflect that two-qubit gates induce more noise than one-qubit gates.\par
The estimated execution efficiency
\begin{equation} \label{eqn:app:tau}
	\tau(\subS, R, \mathcal{U}, s_1, s_2) := \max_T \tau_T(\subS, R, 
	\mathcal{U}, s_1, s_2) \leq 0
\end{equation}
is defined as the largest estimated transpilation efficiency, \cref{eqn:app:tauT}, for all possible transpilation procedures. Summarized, the estimated execution efficiency represents the largest possible sum over the penalty scores of all gates and can therefore be used to estimate how efficiently the corresponding circuit can be executed on the hardware device of interest.\par
More generally, the efficiency of a transpilation procedure may not only depend on the absolute number of gates, but also the noise properties of individual qubits and gates as well as low-level hardware realizations to achieve a better performance \cite{swamit2018,murali2019,wilson2020,nathan2021,mapomatic2022}. We do not consider this more complex case here, which can, however, be brought to the same formalism by defining a suitably modified penalty function.

\begin{figure}[!t]
	\centering
	\begin{subfigure}[t]{.59\linewidth}
		\centering
		\includefigure{1}{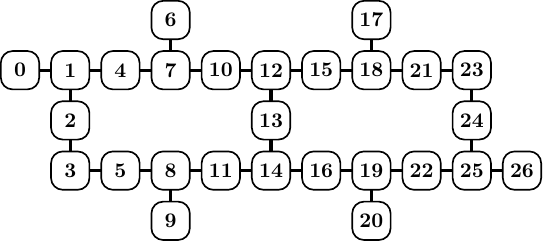}
		\caption{\QCehningen}
		\label{fig:app:backends:ehningen}
	\end{subfigure}%
	\hfill
	\begin{subfigure}[t]{.39\linewidth}
		\centering
		\includefigure{1}{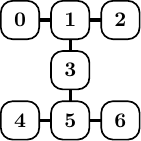}
		\caption{\QCoslo}
		\label{fig:app:backends:oslo}
	\end{subfigure}%	
	\caption{Hardware connectivity map for the two \ac{NISQ} devices of interest. We show the physical qubit indices (unique identifiers) and connections between qubits that allow the realization of joint $\CNOT$ gates.}
	\label{fig:app:backends}
\end{figure}

\appendixsection{NISQ devices} \label{sec:app:backends}
We use two \ac{NISQ} devices to perform our experiments: \QCehningen (accessible via \cite{ibmq2022a}) and \QCoslo (accessible via \cite{ibmq2022b}). The specifications of these devices---as provided by IBM---are listed in \cref{tab:app:backends}. Both \acp{QPU} can realize four elementary gates that form a universal gate set: (i) three one-qubit gates, namely the Z-rotation gate $\RZ(\lambda)$ (parameterized by an angle $\lambda$), the NOT gate $\NOT$, and the SX gate $\SX$, as well as (ii) the two-qubit controlled NOT gate $\CNOT$. In \cref{fig:app:backends}, we show the hardware connectivity maps. Each node represents a qubit labeled with the corresponding physical qubit index (that allows a unique identification) and each edge indicate the respective qubit connections that allow the realization of joint $\CNOT$ gates.
\begin{table}[!t]
	\centering
	\caption{\Ac{NISQ} device specifications as provided by IBM.} \label{tab:app:backends}
	\begin{tabular}{lccccc} \toprule
		\ac{QPU} & processor type & version & \makecell{number \\ of qubits} & \makecell{quantum \\ volume \cite{cross2019}} & CLOPS \cite{wack2021} \\\midrule
		\QCehningen & Falcon r5.11  & 3.1.22 & \num{27} & \num{64} & \num{1.9e3} \\\midrule
		\QCoslo     & Falcon r5.11H & 1.0.14 & \num{7}  & \num{32} & \num{2.6e3} \\\midrule
		\ac{QPU} & \makecell{median \\ $\CNOT$ error} & \makecell{median \\ readout error} & median $T_1$ & median $T_2$ & reference \\\midrule
		\QCehningen & \num{7.218e-3} & \num{9.300e-3} & \SI{127.09}{\micro\second} & \SI{143.27}{\micro\second} & \cite{ibmq2022a} \\\midrule
		\QCoslo     & \num{9.748e-3} & \num{1.950e-2} & \SI{148.03}{\micro\second} & \SI{41.79}{\micro\second}  & \cite{ibmq2022b} \\\bottomrule
	\end{tabular}
\end{table}

\begin{figure}[!t]
	\centering
	\begin{subfigure}[t]{.15\linewidth}
		\centering
		\includefigure{1}{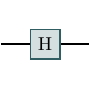}
		\caption{$\HADAMARD$}
		\label{fig:app:circuit-symbols:h}
	\end{subfigure}%
	\hfill
	\begin{subfigure}[t]{.15\linewidth}
		\centering
		\includefigure{1}{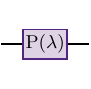}
		\caption{$\PHASE(\lambda)$}
		\label{fig:app:circuit-symbols:p}
	\end{subfigure}%
	\hfill
	\begin{subfigure}[t]{.15\linewidth}
		\centering
		\includefigure{1}{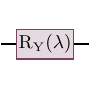}
		\caption{$\RY(\lambda)$}
		\label{fig:app:circuit-symbols:ry}
	\end{subfigure}%
	\hfill
	\begin{subfigure}[t]{.15\linewidth}
		\centering
		\includefigure{1}{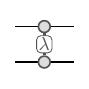}
		\caption{$\CPHASE(\lambda)$}
		\label{fig:app:circuit-symbols:cp}
	\end{subfigure}%
	\hfill
	\begin{subfigure}[t]{.15\linewidth}
		\centering
		\includefigure{1}{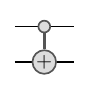}
		\caption{$\CNOT$}
		\label{fig:app:circuit-symbols:cx}
	\end{subfigure}%
	\hfill
	\begin{subfigure}[t]{.15\linewidth}
		\centering
		\includefigure{1}{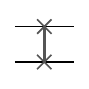}
		\caption{$\SWAP$}
		\label{fig:app:circuit-symbols:swap}
	\end{subfigure}%
	\caption{Overview over the circuit symbols we use throughout the paper: Hadamard gate $\HADAMARD$, phase gate $\PHASE(\lambda)$, Y-rotation gate $\RY(\lambda)$, controlled phase gate $\CPHASE(\lambda)$, controlled NOT gate $\CNOT$, and SWAP gate $\SWAP$ gate. Here, $\lambda$ denotes a real parameter.}
	\label{fig:app:circuit-symbols}
\end{figure}

\appendixsection{Circuit symbols} \label{sec:app:circuit symbols}
Throughout this paper, we use the circuit symbols shown in \cref{fig:app:circuit-symbols} to denote quantum gates. All gates are used as defined by Qiskit \cite{qiskit2021}.

\appendixsection{QSVM marginal contributions} \label{sec:app:qsvm marginal contributions}
In this appendix section, we provide supplementary material for the \ac{QSVM} use case from \cref{sec:quantum support vector machine}. Specifically, we investigate the distribution of marginal contributions $p_i(\Delta_i v)$, \cref{eqn:deltav:p}, for each player index $i$, where $i$ corresponds to the gate index $g$. The results are based on the evaluation from the \QCstatevector simulator with $K=1$. For $r \in \{1,2\}$, we use $\alpha=1$, whereas for $r=3$, we use $\alpha=\num{0.01}$ with a single run. That is, we use the same setup as for \cref{fig:kernel-svm:results:phi}, but consider only the first run for $r=3$.\par
\begin{figure}[!t]
	\centering
	\begin{subfigure}[b]{.32\linewidth}
		\centering
		\includefigure{\scaleA}{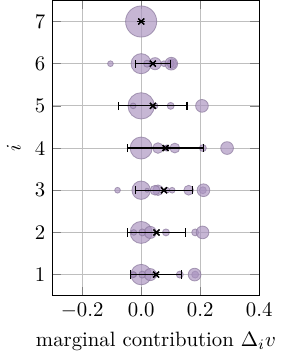}
		\caption{$r=1$}
		\label{fig:app:kernel-svm:results:vdist:1}
	\end{subfigure}%
	\hfill
	\begin{subfigure}[b]{.32\linewidth}
		\centering
		\includefigure{\scaleA}{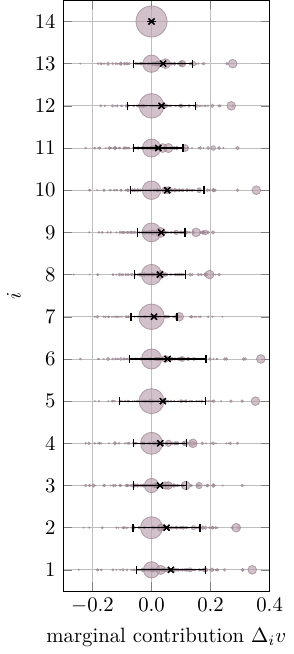}
		\caption{$r=2$}
		\label{fig:app:kernel-svm:results:vdist:2}
	\end{subfigure}%
	\hfill
	\begin{subfigure}[b]{.32\linewidth}
		\centering
		\includefigure{\scaleA}{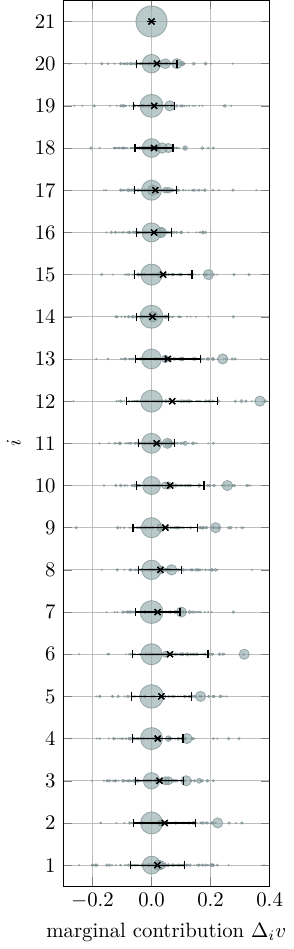}
		\caption{$r=3$}
		\label{fig:app:kernel-svm:results:vdist:3}
	\end{subfigure}%
	\caption{Distribution of marginal contributions $p_i(\Delta_i v)$, \cref{eqn:deltav:p}, for the \ac{QSVM} use case from \cref{sec:quantum support vector machine} with repetition numbers $r \in \{1,2,3\}$. For each player index $i$, we show the marginal contributions as circles with a size that monotonically depends on the corresponding probability. All evaluations are based on  the \QCstatevector simulator with $K=1$. We use $\alpha=1$ for $r \in \{1,2\}$ and $\alpha=\num{0.01}$ for $r=3$. For each player index $i$, we show the mean, \cref{eqn:phi}, and one standard deviation, \cref{eqn:phi:V}.}
	\label{fig:app:kernel-svm:results:vdist}
\end{figure}

The results are shown in \cref{fig:app:kernel-svm:results:vdist}. Each circle represents a marginal contribution with a size that monotonically depends on the corresponding probability (\ie, the larger, the more probable). In addition, we also show the mean, \cref{eqn:phi}, and one standard deviation, \cref{eqn:phi:V}, for each index $i$. For the sake of simplicity, we omit values with $p < \num{0.001}$. Furthermore, we renormalize the probabilities to one for $r=3$ (which are not normalized due to the subsampling).\par
By definition, the mean values from \cref{fig:app:kernel-svm:results:vdist} correspond to the respective \acp{QSV} from \cref{fig:kernel-svm:results:phi}. We find that almost all \acp{QSV} are within one standard deviation of the other \acp{QSV}.

\appendixsection{QGAN timing of experiments} \label{sec:app:qgan timing}
The start and end times for the QGAN experiments from \cref{sec:qgan} are presented in \cref{tab:app:qgan:times}. Here, the configurations and runs refer to the results shown in \cref{fig:qgan:results:phi}.
\begin{table}[!t]
	\centering
	\caption{Timing of the runs from \cref{fig:qgan:results:phi}. For each experiment and each run, we present the start time, the end time and the number of hourly collected error samples between start and end time as described in \cref{sec:qgan}. Each sample consists of a single gate error and a $\CNOT$ error, respectively, for each qubit or qubit pair of interest. The time intervals include waiting times and the execution of \ac{MEM}. All dates are in the format \emph{YYYY-MM-DD HH:MM:SS}.} \label{tab:app:qgan:times}
	\begin{tabular}{lcccc} \toprule
		configuration & run & start time & end time & samples \\\midrule
		\multirow{3}{*}{0-1-2} 
		& 1 & 2022-12-10 16:01:01 & 2022-12-11 19:12:15 & \num{28} \\
		& 2 & 2022-12-11 19:12:15 & 2022-12-12 18:54:38 & \num{24} \\
		& 3 & 2022-12-12 18:54:39 & 2022-12-13 23:51:04 & \num{29} \\
		\multirow{3}{*}{0-1-2 with \ac{MEM}} 
		& 1 & 2022-12-10 16:01:21 & 2022-12-11 20:23:36 & \num{29} \\
		& 2 & 2022-12-11 20:23:36 & 2022-12-13 07:14:06 & \num{35} \\
		& 3 & 2022-12-13 07:14:07 & 2022-12-14 11:24:05 & \num{29} \\
		\multirow{3}{*}{3-5-4} 
		& 1 & 2022-12-21 12:39:12 & 2022-12-23 02:51:03 & \num{39} \\
		& 2 & 2022-12-23 02:51:03 & 2022-12-24 22:27:32 & \num{44} \\
		& 3 & 2022-12-24 22:27:33 & 2022-12-26 05:47:31 & \num{32} \\
		\multirow{3}{*}{3-5-4 with \ac{MEM}} 
		& 1 & 2022-12-21 12:40:53 & 2022-12-23 08:54:24 & \num{45} \\
		& 2 & 2022-12-23 08:54:24 & 2022-12-25 03:22:59 & \num{43} \\
		& 3 & 2022-12-25 03:22:59 & 2022-12-26 07:15:21 & \num{28} \\\bottomrule			
	\end{tabular}
\end{table}

\appendixsection{QAOA optimization results} \label{sec:app:qaoa optimization results}
The optimization results for the max-cut problem from \cref{sec:qaoa} are listed in \cref{tab:app:maxcut-qaoa:opt}. Specifically, we consider three independent runs for different circuit depths $r \in \{1,\dots,7\}$, each with a different random seed for the classical optimizer. The global optimum has a value of \num{9}. 
\begin{table}[!t]
	\centering
	\caption{Optimization goals (\ie, number of cuts in the graph from \cref{fig:maxcut-qaoa:graph}) resulting from \ac{QAOA} for different depths $r$ with three independent runs per depth. The global optimum is highlighted in bold. All calculations are performed on a \QCstatevector simulator.} \label{tab:app:maxcut-qaoa:opt}
	\begin{tabular}{cccc} \toprule
		\multirow{2}{*}{$r$} & \multicolumn{3}{c}{optimization goal} \\ 
		& run 1 & run 2 & run 3 \\ \midrule
		1 & \textbf{9} & \textbf{9} & 6 \\
		2 & \textbf{9} & \textbf{9} & 6 \\
		3 & 6 & \textbf{9} & \textbf{9} \\
		4 & \textbf{9} & \textbf{9} & \textbf{9} \\
		5 & \textbf{9} & \textbf{9} & \textbf{9} \\
		6 & \textbf{9} & \textbf{9} & \textbf{9} \\				
		7 & \textbf{9} & \textbf{9} & \textbf{9} \\\bottomrule		
	\end{tabular}
\end{table}